\documentclass[12pt]{article}
\usepackage{amsmath,amssymb,graphicx,multirow,subfigure,oldgerm,euscript}
\usepackage[section]{placeins}

\begin{document}
\begin{titlepage}
\begin{flushright}
PCCF-RI-02-08 \\
ADP-02-81/T520
\end{flushright}
\renewcommand{\thefootnote}{\fnsymbol{footnote}}
\vspace{-0.5em}
\begin{center}
{\LARGE{ Enhanced direct $CP$ violation in $B^{\pm,0} \rightarrow \pi^{+} \pi^{-} K^{\pm,0}$ }}
\end{center}
\begin{center}
\begin{large}
O. Leitner$^{1,2}$\footnote{oleitner@physics.adelaide.edu.au}, 
X.-H. Guo$^{1}$\footnote{xhguo@physics.adelaide.edu.au},  
A.W. Thomas${^1}$\footnote{athomas@physics.adelaide.edu.au} \\
\end{large}
\vspace{1.5em}
$^1$ Department of Physics and Mathematical Physics, and \\
Special Research Center for the Subatomic Structure of Matter, \\
University of Adelaide, Adelaide 5005, Australia \\
\vspace{0.5em}
$^2$ Laboratoire de Physique Corpusculaire, Universit\'e Blaise Pascal, \\
CNRS/IN2P3, 24 avenue des Landais, 63177 Aubi\`ere Cedex, France 
\end{center}
\vspace{4.5em}
%
\begin{abstract}
\vspace{1.0em}
We investigate in a phenomenological way, direct $CP$ violation in the hadronic decays $B^{\pm,0} \rightarrow
 \pi^{+} \pi^{-} K^{\pm,0}$ where the effect of $\rho - \omega$ mixing is included. If $N_{c}^{eff}$ (the effective 
parameter associated with factorization) is constrained using 
 the most recent   experimental branching ratios (to $\rho^{0}K^{0}, \rho^{\pm}K^{\mp}, \rho^{\pm}K^{0}, \rho^{0}
K^{\pm}$ and $\omega K^{\pm}$) from the BABAR, BELLE and CLEO Collaborations, we get a maximum
  $CP$ violating asymmetry, $a_{max}$, in the range  $-25\%$ to $+49\%$ for $B^{-} \rightarrow 
\pi^{+}\pi^{-} K^{-}$ and
$-24\%$ to $+55\%$  for ${\Bar B}^{0} \rightarrow \pi^{+}\pi^{-} {\Bar K}^{0}$. We also find that $CP$ violation is
  strongly dependent
 on the Cabibbo-Kobayashi-Maskawa matrix elements. Finally, we show that the  sign of 
$\sin \delta$ is always positive in the
allowed range of $N_{c}^{eff}$ and hence, a measurement of direct $CP$ violation in $B^{\pm,0}  
\rightarrow  \pi^{+} \pi^{-}  K^{\pm,0}$ would 
 remove the mod$(\pi)$ ambiguity in ${\rm arg}\left[ - \frac{V_{ts}V_{tb}^{\star}}{V_{us}V_{ub}^{\star}}\right]$. 
\end{abstract}
\vspace{11.5em}
PACS Numbers: 11.30.Er, 12.39.-x, 13.25.Hw.
%
%
\end{titlepage}
\newpage
%
%
%
\section{Introduction}\label{intro}
%
%
The study of $CP$ violation in $B$ decays is one of the most important aims for  the $B$ factories. The 
relative large  $CP$  violating effects expected in $B$ meson decays   should provide
efficient tests of the standard model through the Cabibbo-Kobayashi-Maskawa (CKM) matrix. It is usually assumed
that a nonzero imaginary  phase angle, $\eta$, is responsible for the  $CP$  violating phenomena. This is why, in 
the past 
few years, numerous theoretical studies and  experiments  have been  conducted in the $B$ meson 
system~\cite{ref1,ref2}
 in order to reduce uncertainties in calculations (e.g.  CKM matrix elements, hadronic matrix elements 
and nonfactorizable effects) and increase our understanding of  $CP$  violation within the standard model framework.
\newline
Direct   $CP$  violating asymmetries in $B$ decays  occur  through the interference of at least two amplitudes with
 different  weak  phase $\phi$ {\em and} strong phase $\delta$. In order to extract  the weak phase  
 (which is determined by the CKM matrix elements) through the measurement of a  $CP$  violating asymmetry,
  one must know  the strong phase   $\delta$ and this 
is usually not well determined. In addition, in order to have a large signal, we have to appeal to  some
 phenomenological mechanism to obtain  a large $\delta$. The charge symmetry violating mixing between
 $\rho^{0}$ and $\omega$ can be extremely important in this regard. In particular, it can lead to   a 
 large  $CP$  violation in $B$ decays, such as $B^{\pm,0} \rightarrow \rho^{0}(\omega) K^{\pm,0} \rightarrow \pi^{+}
 \pi^{-}  K^{\pm,0}$, because   the strong phase  passes  through $90^{o}$ at the $\omega$ 
resonance~\cite{ref3,ref4,ref5}. 
\newline
We have collected  the latest data for $b$ to $s$ transitions concentrating on the  CLEO, BABAR and BELLE branching 
ratio results  in our approach. 
The aim of the present
work is multiple. The main one is to constrain the  $CP$  violating calculation in $B^{\pm,0} \rightarrow \rho^{0}(\omega) 
K^{\pm,0}\rightarrow \pi^{+} \pi^{-} K^{\pm,0}$, including   $\rho-\omega$ mixing and  using the most recent 
experimental data for 
$B \rightarrow \rho K$ decays. The second one is to extract consistent constraints for  $B$ decays into 
$\rho(PS)$
where $PS$ can be either $\pi$ or $K$. In order to extract the strong phase $\delta$,
 we  shall use the factorization approach, in which the hadronic matrix elements of operators are saturated by 
vacuum intermediate states. Moreover, we approximate non-factorizable effects by introducing an effective number of 
colors, $N_{c}^{eff}$.
\newline
In this paper  we  investigate five  phenomenological models with different weak form factors and  determine
 the  $CP$  violating asymmetry, $a$, for $B^{\pm,0} \rightarrow \rho^{0}(\omega) K^{\pm,0} \rightarrow \pi^{+} 
\pi^{-}K^{\pm,0} $  in these models. We select models which are consistent with all the data and  determine 
the  allowed range for  $N_{c}^{eff}$ ($0.66(0.61)< N_{c}^{eff}<2.84(2.82)$). Then,  we  study  the sign of 
 $\sin \delta$ 
in this range of $N_{c}^{eff}$  for all these models. We  also discuss the model  dependence 
of our results  in detail.
\newline
The remainder of this paper is organized as it follows. In Section 2, we present the form of the effective 
Hamiltonian which is based on  the operator  product expansion, together with the values of the corresponding 
Wilson coefficients. 
In Section 3, we give the phenomenological 
formalism for the  $CP$  violating asymmetry in decay processes  including $\rho-\omega$ mixing, where all
 aspects of the calculation of direct  $CP$  violation,
the CKM matrix, $\rho-\omega$ mixing, factorization  and form factors are discussed in detail. In Section 4, we list
all the numerical inputs which are needed for  calculating the  asymmetry, $a$, in $B^{\pm,0} \rightarrow \rho^{0}
(\omega) K^{\pm,0}
 \rightarrow \pi^{+}\pi^{-} K^{\pm,0}$. Section 5 is devoted to results and discussions for these  decays.
 In Section 6, we calculate branching ratios for decays such as $B^{\pm,0} \rightarrow 
\rho^{\pm,0} K^{\pm,0}$ and $B^{\pm} \rightarrow \omega
 K^{\pm}$ as well, and present numerical results over the range of $N_{c}^{eff}$ which is allowed by experimental data
  from the CLEO, BABAR, 
and BELLE Collaborations. In the last section, we summarize our results and determine the allowed range of 
$N_{c}^{eff}$ which is consistent with data for both
 $\rho \pi$ and $\rho K$ decays. Uncertainties in our approach and conclusions are also discussed in this 
section.
%
%
\section{The effective Hamiltonian}\label{part1}
%
%
\subsection{Operator product expansion}\label{part1.1}
%
Operator product expansion (OPE)~\cite{ref6} is a useful tool introduced to analyze  the weak interaction 
of quarks. Defining the decay amplitude $A(M \rightarrow F)$ as
\begin{equation}\label{eq1.1}
A(M \rightarrow F) \propto C_{i}(\mu) \langle F | O_{i}(\mu) | M \rangle \ ,
\end{equation}
where $C_{i}(\mu)$ are the Wilson coefficients (see Section~\ref{part1.2}) and $O_{i}(\mu)$  the operators given by
 the OPE, one sees that
 OPE separates the calculation of the amplitude, $A(M \rightarrow F)$,  into two distinct physical regimes. 
One is related to  {\it hard} or 
short-distance physics, represented by $C_{i}(\mu)$ and calculated  by a perturbative approach. The other    is 
the {\it soft} or long-distance regime. This part must be  treated  by   
non-perturbative approaches such as the  $1/N$ expansion~\cite{ref7}, QCD sum rules~\cite{ref8} or
 hadronic sum rules.

\begin{figure}
\centering\includegraphics[height=5.5cm,clip=true]{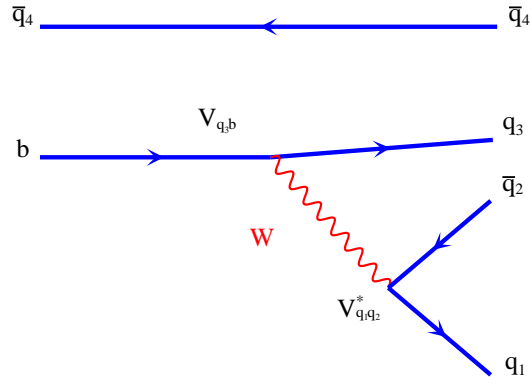}
\caption{Tree diagram, for $B$ decays.}
\label{fig1}
\end{figure}

\begin{figure}
\centering\includegraphics[height=5.5cm,clip=true]{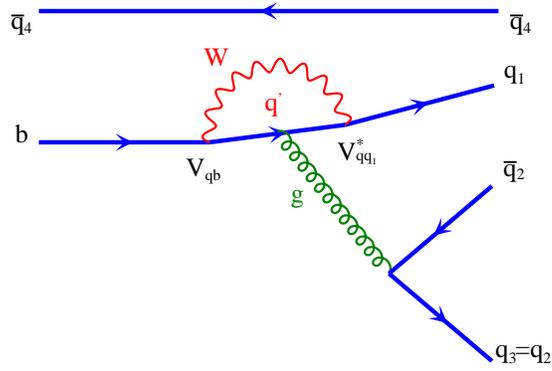}
\caption{QCD penguin diagram, for $B$ decays.}
\label{fig2}
\end{figure}

\begin{figure}
\centering\includegraphics[height=5.5cm,clip=true]{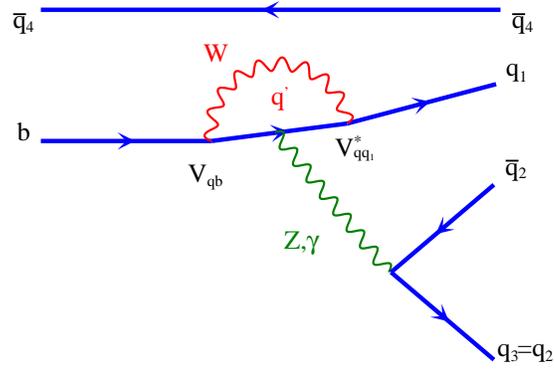}
\caption{Electroweak-penguin diagram, for $B$ decays.}
\label{fig3}
\end{figure}

\begin{figure}
\centering\includegraphics[height=5.5cm,clip=true]{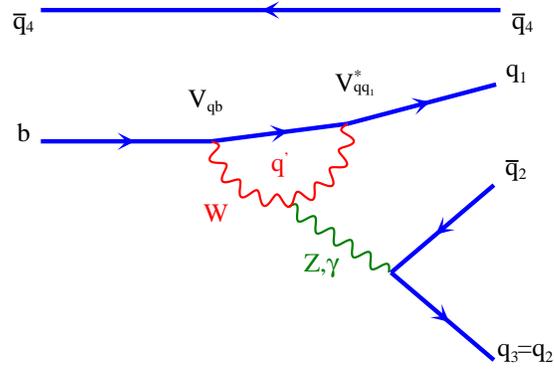}
\caption{Electroweak-penguin diagram (coupling between $Z,\gamma$ and $W$), for $B$ decays.}
\label{fig4}
\end{figure}

\noindent The operators, $O_{i}$,  are  local operators which 
can be written  in the  general  form:
\begin{equation}\label{eq1.2}
O_{n} = ({\bar q}_{i} \Gamma_{n1} q_{j})({\bar q}_{k} \Gamma_{n2} q_{l})\ , 
\end{equation}
where $\Gamma_{n1}$ and $\Gamma_{n2}$ denote
 a combination of gamma matrices and $q$ the quark flavor. They  should respect
 the Dirac structure, the 
color structure and the types of quarks relevant for the  decay being studied. They can be divided into two
classes according to topology:
 tree  operators   ($O_{1}, O_{2}$), and  penguin operators  ($O_{3}$ to $O_{10}$).
For  tree contributions ($W^{\pm}$ is exchanged), the Feynman diagram is shown Fig.~\ref{fig1}. The 
current-current operators related to the tree diagram are the following:
\begin{align}
O_{1}^{s}& = \bar{q}_{\alpha} \gamma_{\mu}(1-\gamma{_5})u_{\beta}\bar{s}_{\beta} \gamma^{\mu}(1-\gamma{_5})
b_{\alpha}\ , \nonumber \\
O_{2}^{s}& = \bar{q} \gamma_{\mu}(1-\gamma{_5})u\bar{s} \gamma^{\mu}(1-\gamma{_5})b\ , 
\end{align}
where $\alpha$ and $\beta$ are the color indices. The  penguin terms  can be divided 
into  two sets. The first  is from the  QCD penguin  diagrams (gluons are exchanged) and the
 second  is from  the 
 electroweak penguin diagrams ($\gamma$ and $Z^{0}$ exchanged). The Feynman  diagram for  the QCD penguin diagram
is shown in
 Fig.~\ref{fig2} and the corresponding operators  are written as follows:
\begin{align}
O_{3}& = \bar{q} \gamma_{\mu}(1-\gamma{_5})b \sum_{q\prime}\bar{q}^{\prime}\gamma^{\mu}(1-\gamma{_5})
q^{\prime}\ , \nonumber \\ 
O_{4}& =\bar{q}_{\alpha} \gamma_{\mu}(1-\gamma{_5})b_{\beta} 
\sum_{q\prime}\bar{q}^{\prime}_{\beta}\gamma^{\mu}(1-\gamma{_5})q^{\prime}_{\alpha}\ , 
\end{align}
\begin{align}
O_{5}& =\bar{q} \gamma_{\mu}(1-\gamma{_5})b \sum_{q'}\bar{q}^
{\prime}\gamma^{\mu}(1+\gamma{_5})q^{\prime}\ , \nonumber \\ 
O_{6}& =\bar{q}_{\alpha} \gamma_{\mu}(1-\gamma{_5})b_{\beta} 
\sum_{q'}\bar{q}^{\prime}_{\beta}\gamma^{\mu}(1+\gamma{_5})q^{\prime}_{\alpha}\ ,   
\end{align}
where $q^{\prime}= u,d,s,c$.
Finally,  the electroweak penguin operators arise from the  two Feynman diagrams represented 
in Fig.~\ref{fig3}
($Z,\gamma$ exchanged from a quark line) and Fig.~\ref{fig4} ($Z,\gamma$ exchanged from the $W$ line). They have the
following expressions:  
\begin{align}
O_{7}& =\frac{3}{2}\bar{q} \gamma_{\mu}(1-\gamma{_5})b \sum_{q'}e_{q^{\prime}}
\bar{q}^{\prime} \gamma^{\mu}(1+\gamma{_5})q^{\prime}\ , \nonumber \\ 
O_{8}& =\frac{3}{2}\bar{q}_{\alpha} \gamma_{\mu}(1-\gamma{_5})b_{\beta} 
\sum_{q'}e_{q^{\prime}}\bar{q}^{\prime}_{\beta}\gamma^{\mu}(1+\gamma{_5})q^{\prime}_{\alpha}\ , \nonumber \\
O_{9}& =\frac{3}{2}\bar{q} \gamma_{\mu}(1-\gamma{_5})b \sum_{q'}e_{q^{\prime}}
\bar{q}^{\prime} \gamma^{\mu}(1-\gamma{_5})q^{\prime}\ , \nonumber \\
 O_{10}& =\frac{3}{2}\bar{q}_{\alpha} \gamma_{\mu}(1-\gamma{_5})b_{\beta} 
\sum_{q'}e_{q^{\prime}}\bar{q}^{\prime}_{\beta}\gamma^{\mu}(1-\gamma{_5})q^{\prime}_{\alpha}\ ,
\end{align}
where  $e_{q^{\prime}}$ denotes the  electric  charge of $q^{\prime}$. 
%
%
\subsection{Wilson coefficients}\label{part1.2}
%
As we  mentioned in the preceding  subsection,  the Wilson 
coefficients~\cite{ref9}, $C_{i}(\mu)$,  represent the 
physical contributions from scales higher than $\mu$ (the OPE describes physics for scales lower than $\mu$).
 Since  QCD has the property of  asymptotic freedom, they can 
be calculated in perturbation theory. The Wilson coefficients include contributions of  all heavy particles,
such as  the top quark, the $W$ bosons, and the charged Higgs boson. Usually, the scale $\mu$ is chosen to be of  
  order  $O(m_{b})$ for  $B$ decays. Wilson coefficients have been  calculated to the next-to-leading order (NLO).
The evolution of $C(\mu)$ (the matrix that includes $C_{i}(\mu)$) is given by,
\begin{equation}\label{eq1.3}
C(\mu)= U(\mu,M_{W})C(M_{W})\ , 
\end{equation}
where $U(\mu,M_{W})$  is the QCD evolution matrix:
\begin{equation}\label{eq1.4}
U(\mu,M_{W})= \biggl[ 1+ \frac{\alpha_{s}(\mu)}{4 \pi}J \biggr] U^{0}(\mu,M_{W}
) \biggl[1- \frac{\alpha_{s}(M_{W})}{4 \pi}J \biggr] \ ,
\end{equation}
with $J$ the matrix  summarizing the next-to-leading order corrections and $U^{0}(\mu,M_{W})$ the evolution matrix
in the leading-logarithm  approximation.
%
%
%
Since the strong interaction is independent of
quark flavor, the $C(\mu)$ are the same for all $B$ decays. At the scale 
$\mu=m_{b}=5$ GeV,  $C(\mu)$ take   the  values summarized in Table~\ref{tab1}~\cite{ref10,ref11}.

To be consistent, the matrix elements of the operators, $O_{i}$, should also be renormalized to the one-loop 
order. This results in the effective Wilson coefficients, $C_{i}^{\prime}$, which  satisfy the constraint,
\begin{eqnarray}\label{eq1.6}
C_{i}(m_{b})\langle O_{i}(m_{b})\rangle=C_{i}^{\prime}{\langle O_{i}\rangle}^{tree}\ , 
\end{eqnarray}
where ${\langle O_{i}\rangle}^{tree}$ are the matrix elements at the tree level. These matrix elements   will be
 evaluated 
in the factorization approach. From  Eq.~(\ref{eq1.6}), the relations between $C_{i}^{\prime}$ and $C_{i}$ 
are~\cite{ref10,ref11},
\begin{align}\label{eq1.7}
C_{1}^{\prime}& =C_{1}\ ,\; \nonumber &
C_{2}^{\prime}& =C_{2}\ , \nonumber \\
C_{3}^{\prime}& =C_{3}-P_{s}/3\ ,\; \nonumber &
C_{4}^{\prime}& =C_{4}+P_{s}\ , \nonumber \\
C_{5}^{\prime}& =C_{5}-P_{s}/3\ ,\; \nonumber &
C_{6}^{\prime}& =C_{6}+P_{s}\ , \nonumber \\
C_{7}^{\prime}& =C_{7}+P_{e}\ ,\; \nonumber &
C_{8}^{\prime}& =C_{8}\ , \nonumber \\
C_{9}^{\prime}& =C_{9}+P_{e}\ ,\;  & 
C_{10}^{\prime}& =C_{10}\ ,
\end{align}
where,
\begin{align}
P_{s} & =(\alpha_{s}/8\pi)C_{2}(10/9+G(m_{c},\mu,q^{2}))\ , \nonumber \\
P_{e} & =(\alpha_{em}/9\pi)(3C_{1}+C_{2})(10/9+G(m_{c},\mu,q^{2}))\ , 
\end{align}
and
\begin{eqnarray}
 G(m_{c},\mu,q^{2})=4\int_{0}^{1}dxx(x-1){\rm ln} \frac{m_{c}^{2}-x(1-x)q^{2}}{\mu^{2}}\ .
\end{eqnarray}
Here $q^{2}$ is   the typical  momentum transfer of the gluon or photon in the penguin diagrams and 
$G(m_{c},\mu,q^{2})$ 
has the following explicit expression~\cite{ref12}, 
\begin{align}\label{eq1.8}
&\Re e\;  G  = \frac{2}{3} \left({\rm ln} \frac{m_{c}^{2}}{\mu^{2}}- \frac{5}{3}-4 \frac{m_{c}^{2}}{q^{2}}+
\left(1+2\frac{m_{c}^{2}}{q^{2}}\right)\sqrt{1-4\frac{m_{c}^{2}}{q^{2}}}{\rm ln} \frac{1+\sqrt{1-4
\frac{m_{c}^{2}}{q^{2}}}}{1-\sqrt{1-4\frac{m_{c}^{2}}{q^{2}}}}\right), \nonumber   \\
&\Im m \; G  = -\frac{2}{3}\left(1+2\frac{m_{c}^{2}}{q^{2}}\right)\sqrt{1-4\frac{m_{c}^{2}}{q^{2}}}\ . 
\end{align}
Based on simple arguments  at the quark level, the value of $q^{2}$ is chosen in the range
 $0.3 < q^{2}/m_{b}^{2} < 0.5$~\cite{ref3,ref4}. From Eqs.~(\ref{eq1.7}-\ref{eq1.8}) 
we can obtain numerical values for  
 $C_{i}^{\prime}$. These values are listed in  Table~\ref{tab2}, where we have 
taken $\alpha_{s}(m_{Z})=0.112, \;\;\;   \alpha_{em}(m_{b})=1/132.2,\;\;\;  m_{b}=5$ GeV,
 and $ \;\; m_{c}=1.35$ GeV.
%
%
\subsection{Effective  Hamiltonian}\label{part1.3}
%
%
In any phenomenological treatment of the weak decays of hadrons, the starting point is the weak effective
Hamiltonian at low energy~\cite{ref13}. It is
obtained by integrating out the heavy fields (e.g. the top quark, $W$ and $Z$ bosons) from the standard model
 Lagrangian. It can be written as,
\vspace{-0.5em}
\begin{equation}\label{eq1.9}
{\cal H}_{eff}=\frac {G_{F}}{\sqrt 2} \sum_{i} V_{CKM} C_{i}(\mu)O_i(\mu)\ ,
\end{equation}
where $G_{F}$ is the Fermi constant, $V_{CKM}$ is the CKM matrix element (see Section~\ref{part2.1}), $C_{i}(\mu)$ 
are
the Wilson coefficients (see Section~\ref{part1.2}), $O_i(\mu)$ are the operators from the operator product 
expansion 
(see Section~\ref{part1.1}), and $\mu$ represents the renormalization scale. We emphasize  that the amplitude 
corresponding
to the effective Hamiltonian for a given decay  is independent of the scale $\mu$. In the present case, since we 
analyze direct  $CP$  violation in 
$B$ decays, we take into account both tree and penguin
 diagrams.
For the penguin diagrams, we include all operators $O_{3}$ to $O_{10}$. Therefore, the effective Hamiltonian used will
 be,
\vspace{-0.5em}
\begin{equation}\label{eq1.10}
{\cal H}_{eff}^{\bigtriangleup B=1}=\frac {G_{F}}{\sqrt 2} \biggl[ V_{ub}V_{us}^{\ast}(C_{1}O_{1}^{s} + 
C_{2}O_{2}^{s})- V_{tb}V_{ts}^{\ast} \sum_{i=3}^{10} C_{i}O_{i} \biggr] + H.c.\ ,
\end{equation}
\vspace{-0.5em}
and consequently, the decay amplitude can be expressed as  follows, 
\begin{multline}\label{eq1.11}
A(B \rightarrow P V) =
\frac {G_{F}}{\sqrt 2} \biggl[  V_{ub}V_{us}^{\ast}\bigl( C_{1}\langle P V | O_{1}^{s}| B \rangle + 
C_{2}\langle P V |O_{2}^{s}| B \rangle \bigr) - \\
 V_{tb}V_{ts}^{\ast} \sum_{i=3}^{10} C_{i}\langle P V |O_{i}| B \rangle \biggr]+ H.c.\ ,
\end{multline}
where $\langle P V |O_{i}| B \rangle$ are the hadronic matrix elements.  They  describe the transition between the 
initial
 state and the final state for scales lower than $\mu$ and include, up to now, the main  uncertainties in the calculation 
since they involve  non-perturbative effects.

%
%
\section{ $CP$  violation in $B^{\pm,0} \rightarrow \rho^{0}(\omega)K^{\pm,0} \rightarrow \pi^{+} \pi^{-}
 K^{\pm,0}$}\label{part2}
%
%
Direct  $CP$  violation in a decay process  requires that the two  $CP$  conjugate decay processes have different
 absolute values for their amplitudes~\cite{ref14}.  Let us  start from the usual 
definition of asymmetry,
\begin{equation}\label{eq1.12}
a(B \rightarrow F) = \frac{ \Gamma(B \rightarrow F) -  \Gamma({\bar B}\rightarrow {\bar F})}
{ \Gamma(B \rightarrow F) +  \Gamma({\bar B} \rightarrow {\bar F})}\ , 
\end{equation}
which gives
\begin{equation}\label{eq1.13}
a(B \rightarrow F) = \frac{ | A(B \rightarrow F)|^{2} - | {\bar A}({\bar B} \rightarrow {\bar F} )|^{2}}
{| A(B \rightarrow F)|^{2} + | {\bar A}({\bar B} \rightarrow {\bar F} )|^{2}}\ , 
\end{equation}
where $A(B \rightarrow F)$ is the amplitude for the considered  decay, which in general can be written as
 $A(B \rightarrow F)= |A_{1}| e^{i \delta_{1} +  i\phi_{1}} + |A_{2}| e^{i \delta_{2} + i\phi_{2}}$. Hence one gets
\begin{equation}\label{eq1.14}
a(B \rightarrow F) = \frac{-2|A_{1}||A_{2}| \sin(\phi_{1}- \phi_{2}) \sin(\delta_{1}-\delta_{2})}
{|A_{1}|^{2}+ 2|A_{1}||A_{2}|\cos(\phi_{1}-\phi_{2})\cos(\delta_{1}-\delta_{2}) + |A_{2}|^{2}}\ .
\end{equation}
Therefore, in order to obtain direct  $CP$  violation, the  $CP$  asymmetry parameter $a$ needs a strong 
phase {\it difference},
$\delta_{1}-\delta_{2}$,
coming from the hadronic matrix {\it and}  a weak phase {\it difference}, $\phi_{1}-\phi_{2}$,  coming from 
the CKM matrix.
%
%
\subsection{CKM matrix}\label{part2.1}
%
%
 In phenomenological applications, the widely used  CKM matrix  parametrization    is
 the {\it Wolfenstein parametrization}~\cite{ref15}.
In this approach, the four independent parameters are $\lambda, A, \rho$ and $ \eta$. Then, by expanding each 
element of the
matrix  as a power series of the parameter $\lambda = \sin \theta_{c} = 0.2209$ ($\theta_{c}$ is
the Gell-Mann-Levy-Cabibbo angle), one gets ($O(\lambda^4)$ is neglected)
\begin{equation}\label{eq1.15}
{\hat V}_{CKM}= \left( \begin{array}{ccc}
1-\frac{1}{2} \lambda^{2} &  \lambda                    & A\lambda^{3}(\rho-i\eta) \\
-\lambda                  & 1-\frac{1}{2}\lambda^{2}    & A\lambda^{2}             \\
A\lambda^{3}(1-\rho-i\eta)& -A\lambda^{2}               &      1                   \\
\end{array}  \right)\ ,
\end{equation}
where $\eta$  plays the role of the  $CP$-violating phase. In this parametrization, even though it is an approximation
in $\lambda$, the CKM matrix satisfies unitarity exactly, which means,
\begin{equation}\label{eq1.16}
{\hat V}_{CKM}^{\dagger} \cdot {\hat V}_{CKM} = {\hat I} = {\hat V}_{CKM} \cdot  {\hat V}_{CKM}^{\dagger}\ .
\end{equation}
%
%
%
\subsection{$\rho-\omega$ mixing}\label{part2.2}
%
%
In the vector meson dominance model~\cite{refa1}, the photon propagator is dressed by coupling to  vector mesons.
 From this, the $\rho-\omega$ mixing mechanism~\cite{refa2}  was developed. 
Let $A$ be the amplitude for the decay $B \rightarrow \rho^{0} ( \omega ) K \rightarrow  \pi^{+}  \pi^{-} K$,
 then one has,
\begin{equation}\label{eq1.17}
A=\langle  K  \pi^{-} \pi^{+}|H^{T}|B \rangle + \langle  K  \pi^{-} \pi^{+}|H^{P}|B  \rangle\ ,
\end{equation}
with $H^{T}$ and $H^{P}$ being the Hamiltonians for the tree and penguin operators. We can 
define the relative magnitude and phases between these two contributions  as follows,
\begin{align}\label{eq1.18}
A &= \langle  K  \pi^{-} \pi^{+}|H^{T}| B \rangle [ 1+re^{i\delta}e^{i\phi}]\ , \nonumber \\   
\bar {A} &= \langle \bar{K}  \pi^{+}  \pi^{-}|H^{T}|\bar {B} \rangle [ 1+re^{i\delta}e^{-i\phi}]\ ,  
\end{align}
where $\delta$ and $\phi$ are strong and weak phases, respectively. The phase $\phi$ arises from the 
appropriate combination of CKM matrix elements, and   
$ \phi={\rm arg}[(V_{tb}V_{ts}^{\star})/(V_{ub}V_{us}^{\star})]$. 
As a result, $\sin \phi$ is equal 
to $\sin \gamma$ with $\gamma$ defined in the standard way~\cite{ref16}. The parameter, $r$, is the 
absolute value of the ratio of tree and penguin amplitudes:
\begin{equation}\label{eq1.19}
r \equiv \left| \frac{\langle \rho^{0}(\omega) K|H^{P}|B \rangle}{\langle\rho^{0}(\omega)
K|H^{T}|B \rangle} \right|.
\end{equation}
In order to obtain a  large signal for direct  $CP$  violation, we need some mechanism to make both 
$\sin\delta$  and  $r$ large. We stress that   $\rho-\omega$ mixing has the dual advantages that the strong 
phase difference is large (passing through $90^{o}$ at the $\omega$ resonance) and well known~\cite{ref4, ref5}.
With this mechanism (see Fig.~\ref{fig5}), to first  order in  isospin violation, we have the following results 
when the invariant 
mass of $\pi^{+}\pi^{-}$ is near the $\omega$ resonance mass,
\begin{align}\label{eq1.20}
\langle K \pi^{-} \pi^{+}|H^{T}|B  \rangle & = \frac{g_{\rho}}{s_{\rho}s_{\omega}}
 \tilde{\Pi}_{\rho \omega}t_{\omega} +\frac{g_{\rho}}{s_{\rho}}t_{\rho}\ , \nonumber \\
\langle  K \pi^{-} \pi^{+}|H^{P}|B  \rangle & = \frac{g_{\rho}}{s_{\rho}s_{\omega}} 
\tilde{\Pi}_{\rho \omega}p_{\omega} +\frac{g_{\rho}}{s_{\rho}}p_{\rho}\ .
\end{align}
Here $t_{V} \; (V=\rho \;{\rm  or} \; \omega) $ is the tree amplitude and $p_{V}$ the penguin amplitude for 
 producing a vector meson, $V$, $g_{\rho}$ is the coupling for $\rho^{0} \rightarrow \pi^{+}\pi^{-}$, 
$\tilde{\Pi}_{\rho \omega}$ is the effective $\rho-\omega$ mixing amplitude, and $s_{V}$  is  from the inverse 
 propagator of the vector meson $V$,
\begin{equation}\label{eq1.21}
s_{V}=s-m_{V}^{2}+im_{V}\Gamma_{V}\ , 
\end{equation}
with $\sqrt s$ being the invariant mass of the $\pi^{+}\pi^{-}$ pair. 
We stress that the direct coupling $ \omega \rightarrow \pi^{+} \pi^{-} $ is effectively absorbed into 
$\tilde{\Pi}_{\rho \omega}$~\cite{ref17},  leading  to the explicit $s$ dependence of $\tilde{\Pi}_{\rho \omega}$. 
Making the expansion  $\tilde{\Pi}_{\rho \omega}(s)=\tilde{\Pi}_{\rho \omega}(m_{\omega}^{2})+(s-m_{w}^{2}) 
\tilde{\Pi}_{\rho \omega}^{\prime}(m_{\omega}^{2})$, the  $\rho-\omega$ mixing parameters were determined in 
the fit of Gardner and O'Connell~\cite{ref18}: $\Re e \; 
\tilde{\Pi}_{\rho \omega}(m_{\omega}^{2})=-3500 \pm 300 \; {\rm MeV}^{2}, \;\;\; \Im m \; \tilde{\Pi}_{\rho \omega}
(m_{\omega}^{2})= -300 \pm 300 \; {\rm MeV}^{2}$, and  $\tilde{\Pi}_{\rho \omega}^{\prime}
(m_{\omega}^{2})=0.03 \pm 0.04$. In practice, the effect of the derivative term is negligible.
%
%
>From Eqs.~(\ref{eq1.17}, \ref{eq1.20}) one has
\vspace{0.5em}
\begin{equation}\label{eq1.22}
 re^{i \delta} e^{i \phi}= \frac{ \tilde {\Pi}_{\rho \omega}p_{\omega}+s_{\omega}p_{\rho}}{\tilde 
{\Pi}_{\rho \omega} t_{\omega} + s_{\omega}t_{\rho}}\ . 
\end{equation}
%
Defining
\vspace{-1.5em}
\begin{center}
\begin{equation}\label{eq1.23}
\frac{p_{\omega}}{t_{\rho}} \equiv r^{\prime}e^{i(\delta_{q}+\phi)}\ , \;\;\;\;
\frac{t_{\omega}}{t_{\rho}} \equiv \alpha e^{i \delta_{\alpha}}\ , \;\;\;\;
\frac{p_{\rho}}{p_{\omega}} \equiv \beta e^{i \delta_{\beta}}\ , 
\end{equation}
\end{center}
where $ \delta_{\alpha}, \delta_{\beta}$ and $ \delta_{q}$ are strong phases (absorptive part).
Substituting Eq.~(\ref{eq1.23}) into  Eq.~(\ref{eq1.22}), one finds:
\vspace{0.5em}
\begin{equation}\label{eq1.24}
re^{i\delta}=r^{\prime}e^{i\delta_{q}} \frac{\tilde{\Pi}_{\rho \omega}+ \beta e^{i \delta_{\beta}} 
s_{\omega}}{s_{\omega}+\tilde{\Pi}_{\rho \omega} \alpha e^{i \delta_{\alpha}}}\ ,
\end{equation}
where
\begin{equation}\label{eq1.25}
 \alpha e^{i \delta_{\alpha}}= f, \;\;\; \beta e^{i \delta_{\beta}}= b+ci, \;\;\;r^{\prime}e^{i\delta_{q}}=d+ei\ ,
\end{equation}
and using Eq.~(\ref{eq1.24}), we obtain the following result when $\sqrt s \sim m_{\omega}$:
\vspace{0.5em}
\begin{equation}\label{eq1.25a}
re^{i \delta}= \frac{C + i D}{(s-m_{\omega}^{2}+ f \Re e \; \tilde{\Pi}_{\rho \omega})^{2}+ (f \Im m \; 
\tilde{\Pi}_{\rho \omega} +m_{\omega} \Gamma_{\omega})^{2}}\ .
\end{equation}
%
Here $C$ and $D$ are defined as:
%
\begin{multline}\label{eq1.25b}
C=\bigl(s-m_{\omega}^{2}+ f \Re e\; \tilde {\Pi}_{\rho \omega}\bigr) \Biggl\{ d\biggl[ \Re e \;\tilde {\Pi}_{ \rho
 \omega} 
+b(s-m_{ \omega}^{2})-cm_{ \omega} \Gamma_{ \omega}\biggr] \\ 
  -e \biggl[ \Im m \;\tilde { \Pi}_{ \rho \omega} +bm_{ \omega} \Gamma_{ \omega}+c(s-m_{ \omega}^{2})\biggr] 
\Biggr\} \\
   + \bigl(f  \Im m\; \tilde { \Pi}_{ \rho \omega}  +m_{ \omega} \Gamma_{ \omega}\bigr) \Biggl\{ e\biggl[
 \Re e \;\tilde 
{\Pi}_{\rho \omega} +b(s-m_{ \omega}^{2})-cm_{ \omega} \Gamma_{ \omega}\biggr]  
 \\
 +d\biggl[ \Im m \;\tilde { \Pi}_{ \rho \omega} 
+bm_{ \omega} \Gamma_{ \omega}+c(s-m_{ \omega}^{2})\biggr] \Biggr\} \ ,
\end{multline}
and
\begin{multline}\label{eq1.26}
 D=\bigl(s-m_{\omega}^{2}+ f \Re e \;\tilde {\Pi}_{\rho \omega}\bigr) \Biggl\{ e \biggl[ \Re e \;\tilde {\Pi}_{ \rho
 \omega} +
d(s-m_{ \omega}^{2})-cm_{ \omega} \Gamma_{ \omega}\biggr] \\ 
  +d \biggl[ \Im m\; \tilde { \Pi}_{ \rho \omega} +bm_{ \omega} \Gamma_{ \omega}+c(s-m_{ \omega}^{2})\biggr] 
 \Biggr\} \\
   - \bigl(f  \Im m \;\tilde { \Pi}_{ \rho \omega}  +m_{ \omega} \Gamma_{ \omega}\bigr) \Biggl\{ d \biggl[
 \Re e \;\tilde 
{\Pi}_{\rho \omega} +b(s-m_{ \omega}^{2})-cm_{ \omega} \Gamma_{ \omega}\biggr]  
\\
  -e\biggl[ \Im m\; \tilde { \Pi}_{ \rho \omega} 
+bm_{ \omega} \Gamma_{ \omega}+c(s-m_{ \omega}^{2})\biggr] \Biggr\}\ . 
\end{multline}
$\alpha e^{i \delta_{\alpha}}$, $\beta e^{i \delta_{\beta}}$, and $r^{\prime}e^{i \delta_{q}}$ will be
 calculated later. In order to get the  $CP$  violating asymmetry, $a$, 
$\sin\phi$ and $\cos\phi$ are needed, where $\phi$ is determined by the CKM matrix elements. 
In the Wolfenstein parametrization~\cite{ref15}, the weak phase comes from 
$[V_{tb} V^{\star}_{ts}/V_{ub} V^{\star}_{us}]$ and one has for the decay $B \rightarrow \rho^{0}(\omega) K$,
\begin{align}\label{eq1.27}
\sin\phi & = \frac{-\eta}{\sqrt {\rho^2+\eta^{2}}}\ , \nonumber \\
\cos\phi & = \frac{-\rho}{\sqrt {\rho^2+\eta^{2}}}\ .
\end{align}
The values used for $\rho$ and $\eta$ will be discussed in Section~\ref{part3.1}.
%
%
\subsection{Factorization}\label{part2.3}
%
With the Hamiltonian given in Eq.~(\ref{eq1.10}) (see Section~\ref{part1.3}), we are ready to evaluate 
the matrix elements  for $B^{\pm,0}\rightarrow 
\rho^{0}(\omega) K^{\pm,0}$.  In the factorization 
approximation~\cite{ref19}, either  $\rho^{0}(\omega)$ or 
$K^{\pm,0}$ is generated by one current which has the appropriate  quantum numbers in the Hamiltonian.
For these  decay processes, two kinds of matrix element products are involved after factorization 
(i.e. omitting Dirac matrices and color labels):  $\langle \rho^{0}(\omega)|(\bar{u}u)|0\rangle 
\langle K^{\pm,0}|(\bar{s}b)|B^{\pm,0}\rangle $ and $ \langle K^{\pm,0}|
(\bar{q}_{1}q_{2})|0\rangle \langle\rho^{0}
(\omega)|(\bar{u}b)|B^{\pm,0}\rangle$, where $q_{1}$ and $q_{2}$ could be $u, s$ or $d$. We will calculate 
them in several phenomenological quark models. 
\newline
The matrix elements for $B \rightarrow X$ and  $B \rightarrow X^{\star}$ (where X and $ X^{\star}$ 
denote pseudoscalar and vector mesons, respectively) can be decomposed as follows~\cite{ref20},
\begin{equation}\label{eq1.28}
\langle X|J_{\mu}|B \rangle =\left( p_{B} + p_{X}- \frac{m_{B}^{2}-m_{X}^{2}}{k^{2}}k \right)_{\mu} 
F_{1}(k^{2})+\frac{m_{B}^{2}-m_{X}^{2}}{k^{2}}k_{\mu}F_{0}(k^{2})\ ,
\end{equation}
and, 
\vspace{-0.8em}
\begin{multline}\label{eq1.29}
\langle X^{\star}|J_{\mu}|B \rangle=\frac{2}{m_{B}+m_{X^{\star}}} \epsilon_{\mu \nu \rho \sigma} 
\epsilon^{\star \nu} p_{B}^{\rho} p_{X^{\star}}^{\sigma}  V(k^{2}) +i \Biggl\{ \epsilon_{\mu}^{\star}(m_{B}
+m_{X^{\star}})A_{1}(k^{2})  \\
- \frac{\epsilon^{\star}  \cdot k}{m_{B}+m_{X^{\star}}} (P_{B}+P_{X^{\star}})_{\mu}A_{2}(k^{2})- \frac 
{\epsilon^{\star}  \cdot k}{k^{2}}2m_{X^{\star}} \cdot k_{\mu}A_{3}(k^{2})
\Biggr\}  \\
+i \frac{\epsilon^{\star} \cdot k}{ k^{2}}2m_{X^{\star}} \cdot k_{\mu}A_{0}(k^{2})\ ,
\end{multline}
where $J_{\mu}$  is the weak current, defined as $J_{\mu}=\bar{q}\gamma^{\mu}(1-\gamma_{5})b \;\;     
  {\rm with} \;\; q=u,d,s$ and  $k=p_{B}-p_{X(X^{\star})}$.  $\epsilon_{\mu}$ is the polarization 
vector of $X^{\star}$.  $F_{0}$ and $ F_{1}$ are the form factors related to the transition $0^{-} 
\rightarrow 0^{-}$, while   $A_{0}, A_{1}, A_{2}, A_{3}$ and $ V$ are the form factors that describe the 
transition $0^{-} \rightarrow 1^{-}$.
Finally, in order to cancel the poles at $q^{2}=0$, the form factors respect the conditions:
\begin{equation}\label{eq1.30}
F_{1}(0) = F_{0}(0), \;\;  A_{3}(0) = A_{0}(0)\ ,
\end{equation}
and they also  satisfy the following relations: 
\begin{gather}\label{eq1.31}
A_{3}(k^{2})  = \frac {m_{B}+m_{X^{\star}}}{2m_{X^{\star}}}A_{1}(k^{2})- \frac {m_{B}-m_{X^{\star}}}
{2m_{X^{\star}}}A_{2}(k^{2})\ . 
\end{gather}

 An argument for factorization  has been given by Bjorken~\cite{ref21}:
the  heavy quark decays are very energetic, so the quark-antiquark pair in a meson in a final state moves 
very fast away from
the localized weak interaction. The hadronization of the quark-antiquark pair occurs far away from the remaining 
quarks.
 Then,
the meson can be factorized out and the interaction between the quark pair in the meson and the remaining quark 
should be  tiny.

In the evaluation of matrix elements, the effective number of colors, $N_{c}^{eff}$, enters through  a Fierz 
transformation. In general, for operator $O_{i}$, one can write,
\begin{equation}\label{eq1.33}
\frac{1}{(N_{c}^{eff})_{i}} = \frac{1}{3} + \xi_{i}\ , {\rm with} \; i=1, \cdots ,10 \ ,
\end{equation}
where $\xi_{i}$ describes non-factorizable effects. We assume $\xi_{i}$ is universal for all the  operators $O_{i}$.
We also ignore the  final state interactions (FSI). After factorization, and   using the decomposition in 
Eqs.~(\ref{eq1.28},~\ref{eq1.29}), one obtains for the process ${\Bar B}^{0} \rightarrow \rho^{0}(\omega){\Bar K}^{0}$,
\begin{equation}\label{eq5.22}
t_{\rho}=  m_{B}|\vec{p}_{\rho}| \bigl( C_{1}^{\prime}+\frac {1}{N_{c}}C_{2}^{\prime}\bigr) 
f_{\rho}F_{1}(m_{\rho}^{2})\ ,
\end{equation}
where $f_{\rho}$ is the $\rho$ decay constant (and to simplify the formulas we use $N_c$ for $N_{c}^{eff}$ in
Eqs. (40)-(50)).
In the same way, we find $t_{\omega}=t_{\rho}$, so that,
\begin{equation}\label{eq5.23}
\alpha e^{i \delta_{\alpha}}=1\ . 
\end{equation}
After calculating the penguin operator contributions, one has,
\begin{equation}\label{eq5.25}
r^{\prime}e^{i \delta_{q} }=-
\frac{p_{\omega}}{(C_{1}^{\prime}+\frac {1}{N_{c}}C_{2}^{\prime})  f_{\rho}F_{1}(m_{\rho}^{2})}\left| 
\frac{V_{tb}V_{ts}^{\star}}{V_{ub}V_{us}^{\star}}\right|\ , 
\end{equation}
and
\begin{multline}\label{eq5.24}
\beta e^{i \delta_{\beta}}=
\frac{ m_{B}| \vec{p}_{\rho}|}{p_{\omega}} \Bigg\{ 
\frac{3}{2}\biggl( (C_{7}^{\prime}+\frac{1}{N_{c}}C_{8}^{\prime})+(C_{9}^{\prime}+\frac{1}{N_{c}}C_{10}^
{\prime}) \biggr) f_{\rho}F_{1}(m_{\rho}^{2}) +  \\
\hspace{4em} \Biggl( (C_{4}^{\prime}+\frac{1}{N_{c}}C_{3}^{\prime})-\frac{1}{2}(C_{10}^{\prime}+
\frac{1}{N_{c}}C_{9}^{\prime}) \\ +
\biggl(-2(C_{6}^{\prime}+\frac{1}{N_{c}}C_{5}^{\prime})+ (C_{8}^{\prime}+\frac{1}{N_{c}}C_{7}^{\prime}) \biggr)
\left[ \frac{m_{K}^{2}}{(m_{b}+m_{d})(m_{d}+m_{s})}\right]  \Biggr) f_{K}A_{0}(m_{K}^{2})]\Bigg\}\ ,
\end{multline}
where  $f_{K}$ is  the $K$ decay constant.
In  Eqs.~(\ref{eq5.25}, \ref{eq5.24}), $p_{\omega}$  has the following form:
\begin{multline}\label{eq5.26}
p_{\omega}= m_{B}|\vec{p}_{\rho}| \Bigg\{
2 \biggl( (C_{3}^{\prime}+\frac {1}{N_{c}}C_{4}^{\prime})+
(C_{5}^{\prime}+\frac {1}{N_{c}}C_{6}^{\prime})\biggr) f_{\rho}F_{1}(m_{\rho}^{2}) \\
\hspace{5.5em} +  \frac{1}{2}\biggl( (C_{7}^{\prime}+\frac {1}{N_{c}}C_{8}^{\prime})+(C_{9}^{\prime}+\frac {1}{N_{c}}
C_{10}^{\prime})\biggr) f_{\rho}F_{1}(m_{\rho}^{2})  \\
+ \biggl((C_{8}^{\prime}+\frac {1}{N_{c}}C_{7}^{\prime})-2 (C_{6}^{\prime}+\frac {1}{N_{c}}C_{5}^{\prime})
 \biggl) \left[ \frac{m_{K}^{2}f_{K}A_{0}(m_{K}^{2})}{(m_{b}+m_{d})(m_{d}+m_{s})}\right]  \\ 
+  \biggl((C_{4}^{\prime}+\frac {1}{N_{c}}C_{3}^{\prime}) -\frac{1}{2} (C_{10}^{\prime}+\frac {1}{N_{c}}C_{9}^{\prime})
\biggr)
f_{K}A_{0}(m_{K}^{2})
\Bigg\}\ ,    
\end{multline}
and the CKM amplitude entering   the $b \rightarrow s$ transition is, 
\vspace{0.5em}
\begin{equation}\label{eq5.27}
\left| \frac{V_{tb}V_{ts}^{\star}}{V_{ub}V_{us}^{\star}}\right|
= \frac{1}{\lambda^{2}} \frac{1}{\sqrt{{\rho}^{2}+{\eta}^{2}}}
= \frac{1}{\lambda^{2}} \frac{1}{|\sin\beta|}\ ,
\end{equation}
with $\beta$ defined as  the unitarity triangle as usual. Similarly, by applying the same formalism, one gets
for the decay $B^{-} \rightarrow \rho^{0}(\omega)K^{-}$,
\vspace{0.5em}
\begin{equation}\label{eq5.22a}
t_{\rho}=  m_{B}|\vec{p}_{\rho}|  \left[ (C_{1}^{\prime}+\frac {1}{N_{c}}C_{2}^{\prime}) 
f_{\rho}F_{1}(m_{\rho}^{2})+ (C_{2}^{\prime}+\frac {1}{N_{c}}C_{1}^{\prime} )  
f_{K}A_{0}(m_{K}^{2}) \right]\ .
\end{equation}
In the same way, we find $t_{\omega}=t_{\rho}$, therefore one has again,
\begin{equation}\label{eq5.23a}
\alpha e^{i \delta_{\alpha}}=1\ . 
\end{equation}
The ratio between penguin and tree operator contributions, which involves CKM matrix elements, is given by,
\vspace{0.5em}
\begin{equation}\label{eq5.25a}
r^{\prime}e^{i \delta_{q} }=-
\frac{p_{\omega}}{(C_{1}^{\prime}+\frac {1}{N_{c}}C_{2}^{\prime})  f_{\rho}F_{1}(m_{\rho}^{2})+ 
(C_{2}^{\prime}+\frac {1}{N_{c}}C_{1}^{\prime} )    f_{K}A_{0}(m_{K}^{2})}\left| 
\frac{V_{tb}V_{ts}^{\star}}{V_{ub}V_{us}^{\star}}\right|\ , 
\end{equation}
and finally,
\begin{multline}\label{eq5.24a}
\beta e^{i \delta_{\beta}}=
\frac{ m_{B}| \vec{p}_{\rho}|}{p_{\omega}} \Bigg\{ (C_{4}^{\prime}+\frac{1}{N_{c}}C_{3}^{\prime})
f_{K}A_{0}(m_{K}^{2}) \\
\hspace{-1em} +\frac{3}{2}\biggl((C_{7}^{\prime}+\frac{1}{N_{c}}C_{8}^{\prime})+(C_{9}^{\prime}+\frac{1}{N_{c}}
C_{10}^{\prime})\biggr)f_{\rho}F_{1}(m
_{\rho}^{2}) + (C_{10}^{\prime}+\frac{1}{N_{c}}C_{9}^{\prime})f_{K}A_{0}(m_{K}^{2}) \\
-2 \biggl((C_{6}^{\prime}+\frac{1}{N_{c}}C_{5}^{\prime})+(C_{8}^{\prime}+\frac{1}{N_{c}}C_{7}^{\prime}) \biggr)
\left[ \frac{m_{K}^{2}f_{K}A_{0}(m_{K}^{2})}{(m_{u}+m_{s})(m_{b}+m_{u})}\right] 
\Bigg\}\ ,
\end{multline}
where  the $\omega$ penguin contribution, $p_{\omega}$,  is:
\begin{multline}\label{eq5.26a}
p_{\omega}= m_{B}|\vec{p}_{\rho}| \Bigg\{ 2 \biggl( (C_{3}^{\prime}+\frac {1}{N_{c}}C_{4}^{\prime})+
(C_{5}^{\prime}+\frac {1}{N_{c}}C_{6}^{\prime})\biggr)f_{\rho}F_{1}(m_{\rho}^{2}) \\
+  \frac{1}{2}\biggl((C_{7}^{\prime}+\frac {1}{N_{c}}C_{8}^{\prime})+(C_{9}^{\prime}+\frac {1}{N_{c}}
C_{10}^{\prime})\biggr) f_{\rho}F_{1}(m_{\rho}^{2}) 
+ \biggl( (C_{4}^{\prime}+\frac {1}{N_{c}}C_{3}^{\prime})+ (C_{10}^{\prime}+\frac {1}{N_{c}}C_{9}^{\prime})\biggr) 
f_{K}A_{0}(m_{K}^{2}) \\
- 2 \biggl((C_{8}^{\prime}+\frac {1}{N_{c}}C_{7}^{\prime})+(C_{6}^{\prime}+\frac {1}{N_{c}}C_{5}^{\prime})
 \biggr) \left[ \frac{m_{K}^{2}}{(m_{u}+m_{s})(m_{b}+m_{u})}\right] f_{K}A_{0}(m_{K}^{2})
 \Bigg\}\ .    
\end{multline}
%
%
\subsection{Form factors}\label{part2.4}
%
The form factors  $F_{i}(k^{2})$ and $A_{j}(k^{2})$  depend on the inner structure of  the 
hadrons. We will adopt here three different theoretical approaches. The first  was proposed  by Bauer, Stech,
and Wirbel~\cite{ref20} (BSW), who used the overlap integrals of wave functions in order to evaluate 
the meson-meson matrix elements of 
the corresponding  current. The momentum  dependence of the form factors is based on a single-pole ansatz.
The second one was  developed by Guo and Huang (GH)~\cite{ref22}. They modified the BSW model by using some 
wave functions 
described in the light-cone framework. 
The last model was given by Ball~\cite{ref23} and Ball and Braun~\cite{ref24}. In this 
case, the form factors are  calculated from QCD sum rules on the light-cone and  leading twist contributions,
radiative corrections, and $SU(3)$-breaking effects are included. 
Nevertheless, all these models use phenomenological form factors which are parametrized by making 
the nearest pole dominance assumption. The explicit $k^{2}$ dependence of the form factor is
as~\cite{ref20,ref22,ref23,
ref24,ref25}:
\begin{gather}\label{eq1.34}
F_{1}(k^{2})=\frac{h_{1}}{\left( 1-\frac{k^{2}}{m_{1}^{2}} \right)^{n}}\ , \;\;\;  \;\;\; 
A_{0}(k^{2})=\frac{h_{A_{0}}}{\left( 1-\frac{k^{2}}{m_{A_{0}}^{2}}\right)^{n}}\ , \nonumber \\
{\rm or} \hspace{37.8em} \nonumber \\
F_{1}(k^{2})=\frac{h_{1}}{1-d_{1}\frac{k^{2}}{m_{B}^{2}}+b_{1}\left( \frac{k^{2}}{m_{B}^{2}}\right)^{2}}\ ,
 \;\;\;  \;\;\; 
A_{0}(k^{2})=\frac{h_{A_{0}}}{1-d_{0}\frac{k^{2}}{m_{B}^{2}}+b_{0}\left( \frac{k^{2}}{m_{B}^{2}}\right)^{2}}\ ,
\end{gather} 
where $n=1,2$, $m_{A_{0}}$ and $m_{1}$ are the pole masses associated with  the transition current, 
 $h_{1}$ and $h_{A_{0}}$
are the values of form factors at $q^{2}=0$, and $d_{i}$ and $b_{i}$ ($i=0,1$)  are  parameters in the  model of 
Ball.
%
%
\section{Numerical inputs}\label{part3}
%
\subsection{CKM values}\label{part3.1}
%
In our numerical calculations  we have several parameters: $q^{2}, N_{c}^{eff}$, and the CKM matrix elements 
in the Wolfenstein parametrization. As mentioned in Section~\ref{part1.2}, the value of $q^{2}$ is conventionally
chosen to be in the range $0.3<q^{2}/{m_{b}}^{2}<0.5$. The CKM matrix, which should be determined from
 experimental data,  is expressed  in terms  of the Wolfenstein parameters, $ A,\; \lambda,\; 
\rho$, and $\eta $~\cite{ref15}.
Here, we shall use  the latest  values~\cite{ref26}  which were  extracted from
charmless semileptonic $B$ decays, ($|V_{ub}|$),  charm semileptonic $B$ decays,  ($|V_{cb}|$),
$s$ and $d$ mass oscillations, $\Delta m_{s}, \Delta m_{d}$,
and  $CP$  violation in the kaon system ($\epsilon_{K}$), ($\rho, \eta$). Hence, one has,
\begin{equation}\label{eq1.35}
\lambda=0.2237\ , \;\; A=0.8113\ , \;\;  0.190 < \rho < 0.268\ , \;\; 0.284< \eta <0.366\ .
\end{equation}
These values  respect the unitarity triangle as well (see also Table~\ref{tab4}).
%
%
\subsection{Quark masses}\label{part3.2}
%
%
The running quark masses are used in order to calculate the 
matrix elements of penguin operators. The quark mass is  taken at the scale $\mu \simeq m_{b}$ in 
$B$ decays. Therefore one has~\cite{ref27},
\begin{align}\label{eq1.36}
m_{u}(\mu=m_{b})& = 2.3 \;{\rm MeV}\ , & m_{d}(\mu=m_{b})& = 4.6 \;{\rm MeV}\ , \nonumber\\
m_{s}(\mu=m_{b})& = 90 \;{\rm MeV}\ , &  m_{b}(\mu=m_{b})& = 4.9 \;{\rm GeV}\ ,
\end{align}
which corresponds to $m_{s}(\mu= 1\;{\rm GeV}) = 140 \;{\rm MeV}$. For meson  masses,
we shall use the following values~\cite{ref16}:
\begin{align}\label{eq1.37}
m_{B^{\pm}}& = 5.279 \; {\rm GeV}\ ,  & m_{B^{0}}&   = 5.279 \; {\rm GeV}\ , \nonumber\\
m_{K^{\pm}}& = 0.493 \;{\rm GeV}\ ,   & m_{K^{0}}&   = 0.497 \;{\rm GeV}\ ,  \nonumber\\
m_{\pi^{\pm}}& = 0.139 \;{\rm GeV}\ , & m_{\pi^{0}} & = 0.135 \;{\rm GeV}\ , \nonumber\\ 
 m_{\rho^{0}}& = 0.769  \;{\rm GeV}\ , &  m_{\omega}& = 0.782 \;{\rm GeV}\ .  
\end{align}
%
%
\subsection{Form factors and decay constants}\label{part3.3}
%
%
In  Table~\ref{tab5} we  list the relevant form factor values at zero 
momentum transfer~\cite{ref20,ref22,ref23,ref24,ref28}  for the
$B \rightarrow K$ and  $B \rightarrow \rho$ transitions.
The different models are defined as follows: models (1) and (3) are the BSW model where the $q^{2}$ dependence of the 
form
factors is described  by  a single- and  a double-pole ansatz, respectively. Models (2) 
and (4)  are the GH model with the 
same momentum dependence as models (1) and (3). Finally,  model (5) refers to the Ball model.
We define the decay constants for  pseudo-scalar ($f_{P}$) and vector ($f_{V}$) mesons as usual by,
\begin{align}\label{eq1.38}
\langle P(q) | \bar{q}_{1} \gamma_{\mu} \gamma_{5} q_{2}| 0 \rangle & = i f_{P} q_{\mu}\ , \nonumber \\
\sqrt{2} \langle V(q) | \bar{q}_{1} \gamma_{\mu} q_{2} | 0 \rangle & = f_{V} m_{V} \epsilon_{V}\ ,
\end{align}
with $q_{\mu}$ being the momentum of  the pseudo-scalar meson, $m_{V}$ and $\epsilon_{V}$ being the mass
and polarization vector of the vector meson, respectively.
 Numerically, in our calculations, we  take~\cite{ref16},
\begin{gather}\label{eq1.39}
   f_{K} = 160 \; {\rm MeV}\ ,  \; 
f_{\rho} \simeq f_{\omega}  = 221 \; {\rm MeV}\ . 
\end{gather}
The  $\rho$ and $\omega$ decay constants are very close and  for simplification (without
any consequences for  results) we choose $f_{\rho} = f_{\omega}$.
%
%
\section{Results and discussion}\label{part4}
%
%
We have investigated the  $CP$  violating asymmetry, $a$, for the two $B$ decays: ${\Bar B}^{0} \rightarrow \rho^{0} {\Bar K}
^{0} \rightarrow
\pi^{+} \pi^{-} {\Bar K}^{0}$ and $B^{-} \rightarrow \rho^{0} K^{-} \rightarrow \pi^{+} \pi^{-} K^{-}$. The results 
are shown in Figs.~\ref{fig6}  and~\ref{fig7} for ${\Bar B}^{0} \rightarrow \pi^{+} \pi^{-}
 {\Bar K}^{0}$, ($a 
=[\Gamma({\Bar B}^{0} \rightarrow \pi^{+} \pi^{-} {\Bar K}^{0}) -\Gamma(B^{0} \rightarrow \pi^{-} \pi^{+}  K^{0})]
/[\Gamma({\Bar B}^{0}\rightarrow \pi^{+} \pi^{-} {\Bar K}^{0}) +\Gamma(B^{0} \rightarrow \pi^{-} \pi^{+}  K^{0})]$), 
where $k^{2}/m_{b}^{2}=0.3(0.5)$ 
and for $N_{c}^{eff}$ equal to $0.61, 0.66, 2.65, 2.69, 2.82$ and $2.84$. Similarly,  in Figs.~\ref{fig8}
 and~\ref{fig9}, the  $CP$  violating asymmetry, $a$, ($=[\Gamma(B^{-}\rightarrow \pi^{+} \pi^{-}K^{-} ) -\Gamma(B^{+}
\rightarrow \pi^{-} \pi^{+}K^{+})]/[\Gamma(B^{-}\rightarrow \pi^{+} \pi^{-}K^{-}) +\Gamma(B^{+}
\rightarrow \pi^{-} \pi^{+}K^{+})]$),
is plotted for $B^{-} \rightarrow \pi^{+} \pi^{-} K^{-}$, where $k^{2}/m_{b}^{2}=0.3(0.5)$ and 
for the same values of $N_{c}^{eff}$ previously applied for ${\Bar B}^{0} \rightarrow
\pi^{+} \pi^{-} {\Bar K}^{0}$. In our numerical calculations, we found that
the  $CP$  violating parameter, $a$, reaches a maximum value, $a_{max}$, when the invariant mass of the $\pi^{+} 
\pi^{-}$ is in the vicinity of the $\omega$ resonance, for a fixed value of $N_{c}^{eff}$. We have studied the model
dependence of $a$ with five models where different form factors have been applied. Numerical results for 
${\Bar B}^{0} \rightarrow \pi^{+} \pi^{-} {\Bar K}^{0}$ and $B^{-} \rightarrow \pi^{+} \pi^{-} K^{-}$ are listed 
in Tables~\ref{tab7}
and~\ref{tab8}, respectively. It appears that the form factor dependence of $a$ for all  models, and in both 
decays, is weaker than the $N_{c}^{eff}$ dependence.

For ${\Bar B}^{0} \rightarrow \pi^{+} \pi^{-} {\Bar K}^{0}$, we have determined the range of  the 
 maximum asymmetry parameter, $a_{max}$, 
when $N_{c}^{eff}$ varies  between $0.66(0.61)$ and $2.84(2.82)$, in the case of $k^{2}/m_{b}^{2}=0.3(0.5)$. 
The evaluation of $a_{max}$ gives allowed values from $37\%(55\%)$ to $-20\%(-24\%)$ for the range of $N_{c}^{eff}$
and CKM matrix elements indicated before. The sign of $a_{max}$ stays positive until $N_{c}^{eff}$ reaches $2.7$. If 
we look at the numerical results for the asymmetries (Table~\ref{tab7}),
for $N_{cmin}^{eff}=0.66(0.61)$ and $k^{2}/m_{b}^{2}=0.3(0.5)$, we find good  agreement between all the models,
 with a maximum  asymmetry, $a_{max}$, around $33\%(45.6\%)$ for the set 
$(\rho_{max},\eta_{max})$, 
and around $26\%(33.2\%)$ for the set $(\rho_{min},\eta_{min})$. The ratio between asymmetries associated with 
the upper and lower limits of $(\rho,\eta)$ is around $1.26(1.37)$. If we consider the maximum asymmetry 
parameter, $a_{max}$, for $N_{cmax}^{eff}=2.84(2.82)$, we  observe a distinction between the   models. Indeed, two
classes of models appear: models $(2)$ and $(4)$ and models $(1,3)$ and $(5)$. For  models $(2)$ and $(4)$, one has
an asymmetry,  $a_{max}$, around $-6\%(-7\%)$ and around $-9\%(-10\%)$  for the upper and lower set of $(\rho,\eta)$,
respectively. The ratio between them is around $1.50(1.42)$. For models $(1,3)$ and $(5)$, the maximum asymmetry is of 
order  $-14.3\%(-16.3\%)$ for $(\rho_{max},\eta_{max})$ and around  $-19.3\%(-23.0\%)$ for $(\rho_{min},\eta_{min})$.
In this  case, the ratio between asymmetries  is around $1.34(1.41)$.

The first reason why the maximum asymmetry, $a_{max}$, can vary so much comes from the element $V_{ub}$.
 The other CKM matrix elements $V_{tb}, V_{ts}$ and $V_{us}$, all proportional to $A$ and 
$\lambda$, are very well measured experimentally  and thus do not interfere in our  results. Only 
 $V_{ub}$, which contains
the $\rho$ and $\eta$ parameters, provides large uncertainties, and thus, large variations for the maximum  
asymmetry. The second reason is  the non-factorizable effects in the transition $b \rightarrow s$. It is 
well known that decays including a $K$ meson (and therefore an $s$ quark) carry more uncertainties than
those involving only a $\pi$ meson ($u$,$d$ quarks). If we look at the asymmetries at $N_{cmin}^{eff}$, all 
 models give almost the same
values, whereas at $N_{cmax}^{eff}$, we obtain different asymmetry values (with, moreover, a change of sign for the
 $CP$  violating asymmetry). 
The  $CP$  asymmetry parameter is more  sensitive to form factors at high values 
of $N_{c}^{eff}$ than at low values of $N_{c}^{eff}$. It appears therefore  that all of the  models 
  investigated  can be divided in two classes, referring to 
the two classes of form factors.

For $B^{-} \rightarrow \pi^{+} \pi^{-} K^{-}$, we have similarly investigated  the 
 $CP$  violating asymmetry. The values of maximum asymmetry parameter, $a_{max}$, for a range of $N_{c}^{eff}$ from 
$0.66(0.61)$ to $2.84(2.82)$, where $k^{2}/m_{b}^{2}=0.3(0.5)$ and for the five models analyzed, are given in
Table~\ref{tab8}. We found that for  this  decay, the  $CP$  violating parameter, $a$, takes values around 
$49\%(46\%)$ to $-22\%(-25\%)$ for the limiting CKM matrix values of $\rho$ and $\eta$ defined before. Once again,
the sign of the asymmetry parameter, $a$, is positive if the value of $N_{c}^{eff}$ stays below $2.7$. If we focus
on $N_{cmin}^{eff}$ equal to $0.66(0.61)$, models $(1,2,3,4)$ and $(5)$ give almost 
the same value which is around $46.6\%(43.6\%)$
for the maximum values of the CKM matrix elements. For the set $(\rho_{min},\eta_{min})$, the maximum asymmetry, $a$,
is around $34.0\%(33.8\%)$. The ratio  between asymmetry values taken at upper and lower limiting $\rho$ and $\eta$
values is around $1.37(1.28)$. Let us have a look at the  $CP$  asymmetry values at $N_{cmax}^{eff}$. As we observed 
for the decay ${\Bar B}^{0} \rightarrow \pi^{+} \pi^{-} {\Bar K}^{0}$, all  models are separated into two distinct 
classes related
to their form factors. For models $(1,3)$ and $(5)$, the value of maximum asymmetry, $a_{max}$, is  around
$-15.6\%(-17.6\%)$ and around $-21\%(-23.6\%)$ for the maximum and minimum values of set $(\rho,\eta)$, respectively.
The calculated ratio  is around
 $1.34(1.34)$, between these two  asymmetries. As regards models $(2)$ and $(4)$, for the same
set of $(\rho,\eta)$, one gets $-11.5\%(-13\%)$ and $-17\%(-18\%)$. In this case, one has $1.47(1.38)$ for the 
 ratio. The reasons for  the differences between the maximum asymmetry parameter, $a_{max}$, are the 
same as in the decay  ${\Bar B}^{0} \rightarrow \pi^{+} \pi^{-} {\Bar K}^{0}$.

By analyzing the $B$ decays, such as ${\Bar B}^{0} \rightarrow \pi^{+} \pi^{-} {\Bar K}^{0}$ and $B^{-} 
\rightarrow \pi^{+} \pi^{-} 
K^{-}$, we found that the  $CP$  violating asymmetry, $a$, depends on the CKM matrix elements, form factors and the
effective parameter $N_{c}^{eff}$ (in order of increasing dependence). As regards the 
CKM matrix elements, the dependence
through the element, $V_{ub}$, contributes to the asymmetry in the ratio between the $\omega$ penguin contributions
 and the $\rho$
tree contributions. It
 also appears that for the upper limit of set $(\rho,\eta)$, we get the higher value asymmetry, $a$,
and vice versa. With regard to the 
 form factors, the dependence at low values of $N_{c}^{eff}$ is very weak although the huge
difference between the phenomenological form factors (models $(2)$ and $(4)$ and models $(1,3)$ and $(5)$) applied
in our calculations. At high values of $N_{c}^{eff}$, the dependence becomes strong and then, the 
 asymmetry appears very
sensitive to form factors. For the effective parameter, $N_{c}^{eff}$, (related to hadronic non-factorizable 
 effects), our results
show explicitly the dependence of the asymmetry parameter  on it. Because of the energy carried by the quark $s$,
intermediate states and final state interactions are not well taken
 into account and may explain this strong sensitivity.
Finally, results obtained at $k^{2}/m_{b}^{2}=0.3(0.5)$, also show  renormalization effects of the Wilson coefficients 
involved
in the weak effective hadronic Hamiltonian. For the ratio between  asymmetries, results give an average value
of order $1.36(1.40)$ for ${\Bar B}^{0} \rightarrow \pi^{+} \pi^{-} {\Bar K}^{0}$ and $1.39(1.33)$ for $B^{-} 
\rightarrow \pi^{+} 
\pi^{-} K^{-}$. This ratio is mainly governed by the term $1/\sin \beta$, where the values of the angles
 $\alpha, \beta$ and $\gamma$
are listed in Table~\ref{tab4}.

As a first conclusion on these numerical results, it is obvious that the 
 dependence of the asymmetry on the effective parameter $N_{c}^{eff}$ is dramatic and therefore it is absolutely
necessary to more efficiently constrain
  its value, in order to use asymmetry, $a$, to determine the CKM parameters $\rho$ and $\eta$. We know that the 
effects of 
$\rho-\omega$ mixing only exist around  $\omega$ resonance. Nevertheless, in Figs.~\ref{fig6},~\ref{fig7},~\ref{fig8},
 and~\ref{fig9}, at small values of $N_{c}^{eff}$, e.g.
 $\simeq 0.6$, the curves show large asymmetry values far away from  
$\omega$ resonance, which is {\it a priori} unexpected. In fact, if we assume that nonfactorizable effects are not
 as important as factorizable contributions, then $N_{c}^{eff}$ should be much bigger (see Eq.~(\ref{eq1.33})).
>From previous analysis on some other $B$ decays such as $B \rightarrow D \pi, B 
\rightarrow \omega \pi$, and  $B \rightarrow \omega K$, it was found that  $N_{c}^{eff}$ should be 
around $2$~\cite{ref28a}. 
Therefore, 
although small values of $N_{c}^{eff}$ are allowed by the experimental data we are considering  in this paper, 
we expect that the value of $N_{c}^{eff}$ 
cannot be so small with more accurate data. We have checked that when $N_{c}^{eff}$ is larger than $1$ the large  $CP$  
asymmetries are confined in the $\omega$ resonance region. With a very small value of $N_{c}^{eff}$, nonfactorizable 
effects have
been overestimated. This means that soft gluon exchanges between $\rho^{0}(\omega)$ and $K$ may affect $\rho-\omega$
mixing and hence lead to the large  $CP$  asymmetries in a region far away from $\omega$ resonance. However, when $\sqrt s$
is  very far from $\omega$ resonance, the  $CP$  asymmetries go to zero as expected.

In spite of the uncertainties discussed previously, the main effect of $\rho-\omega$ mixing in  $B \rightarrow 
\pi^{+} \pi^{-} K$ is the removal of the  ambiguity  concerning the strong phase, $\sin \delta$. In the
 $b \rightarrow s$ transition,
the weak phase in the rate asymmetry is proportional to $\sin \gamma$ where $\gamma = arg[-(V_{ts}V_{tb}^{\star})
/(V_{us}V_{ub}^{\star})]$. Knowing the sign of $\sin \delta$, we are then able to determine the sign  of  $\sin \gamma$
from a measurement of the asymmetry, $a$. In Figs.~\ref{fig10} and~\ref{fig11}, the value of $\sin \delta$ is plotted
as a function of $N_{c}^{eff}$ for  ${\Bar B}^{0} \rightarrow \pi^{+} \pi^{-} {\Bar K}^{0}$ and $B^{-} 
\rightarrow \pi^{+} \pi^{-} 
K^{-}$, respectively. It appears, in both cases, when $\rho-\omega$ mixing mechanism is included, that the sign
of $\sin \delta$ is  positive, for all models studied, until $N_{c}^{eff}$ reaches $2.69(2.65)$ 
 for both $B^{-} \rightarrow \pi^{+} \pi^{-} K^{-}$ and ${\Bar B}^{0} \rightarrow \pi^{+} \pi^{-} 
{\Bar K}^{0}$, when 
$k^{2}/m_{b}^{2}=0.3(0.5)$. For values of $N_{c}^{eff}$ bigger than this limit,  $\sin \delta$ becomes negative.
At the same time, the sign of the 
asymmetry also changes. In Figs.~\ref{fig12}b and~\ref{fig13}b, the ratio of penguin to
tree amplitudes is shown for $B^{\pm,0} \rightarrow \pi^{+} \pi^{-} K^{\pm,0}$, in the  case of $\tilde{\Pi}_{\rho
\omega}=(-3500,-300)$. The critical point around $N_{c}^{eff}=2.7$,  refers to the change of sign
of $\sin \delta$. Clearly, we can use a measurement of the asymmetry, $a$, to eliminate 
the uncertainty mod$(\pi)$ which
is usually involved in the determination of $\gamma$  (through $\sin 2\gamma$).
 If we do not take into account $\rho-\omega$ 
mixing, the  $CP$  violating asymmetry, $a$, remains very small (just a few percent) in both decays. In Figs.~\ref{fig10} 
and~\ref{fig11} (for the evolution of $\sin \delta $) and in Figs.~\ref{fig12}a and~\ref{fig13}a (for the evolution
of penguin to tree amplitudes), for $B^{\pm,0} \rightarrow \pi^{+} \pi^{-} K^{\pm,0}$, we  plot $\sin \delta$ and 
$r$ when $\tilde{\Pi}_{\rho \omega}=(0,0)$ --i.e. without $\rho-\omega$ mixing. There is  a critical point at 
$N_{c}^{eff}=1$ (for ${\Bar B}^{0} \rightarrow \pi^{+} \pi^{-} {\Bar K}^{0}$) and $N_{c}^{eff}=0.24$ 
(for $B^{-} \rightarrow
 \pi^{+} \pi^{-} K^{-}$) for which the value of $\sin \delta$ is at its maximum and   corresponds 
(for the same value of $N_{c}^{eff}$), to  the lowest value of $r$. The last results show the double effect of the
$\rho-\omega$ mixing: the  $CP$  violating asymmetry increases  and the   sign of the strong phase $\delta$ is determined.
%
%
\section{Branching ratios for  $B^{\pm,0} \rightarrow \rho^{0}K^{\pm,0}$}\label{part5}
%
%
\subsection{Formalism}\label{part5.1}
\begin{figure}
\centering\includegraphics[height=7cm,clip=true]{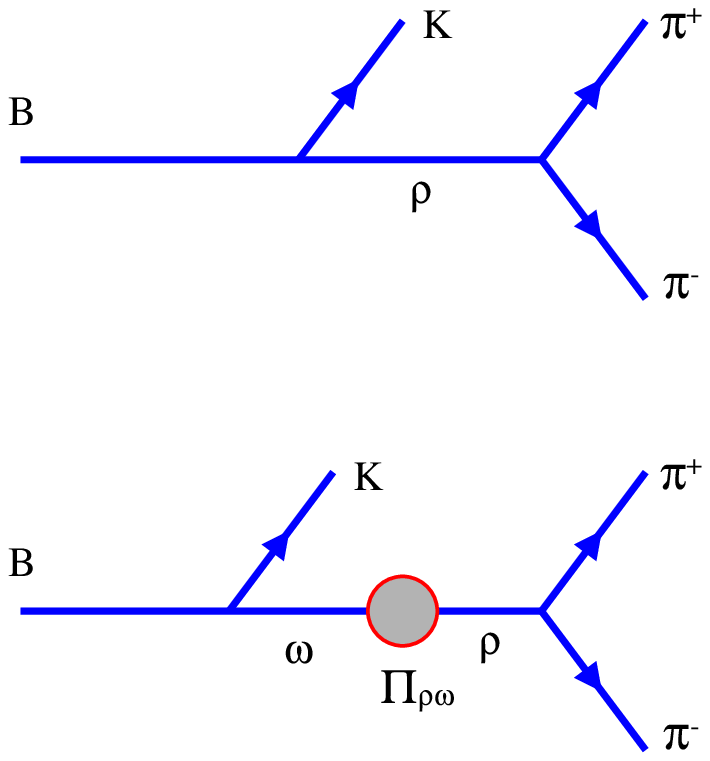}
\caption{$B$ decays without (upper) and with (lower) $\rho - \omega$ mixing.}
\label{fig5}
\end{figure}

\begin{figure}
\centering\includegraphics[height=7cm,clip=true]{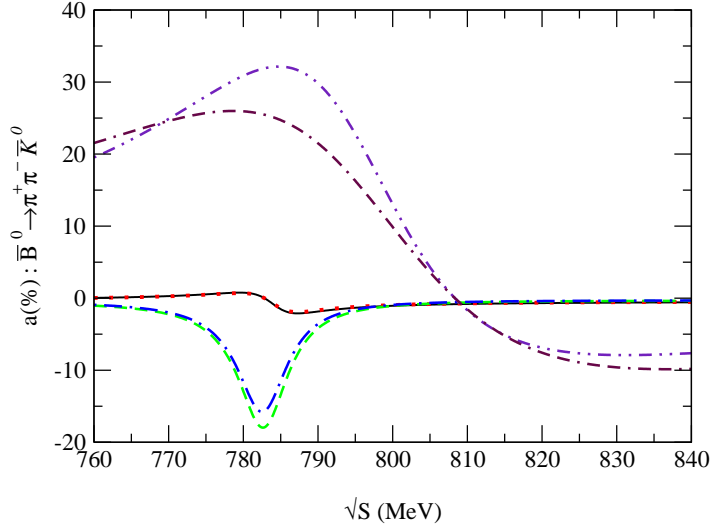}
\caption{ $CP$  violating asymmetry, $a$, for ${\bar B}^{0} \rightarrow \pi^{+} \pi^{-}  {\bar K}^{0}$,
for $k^{2}/m_{b}^{2}=0.3$, for $N_{c}^{eff}= 0.66,2.69,2.84$  and for limiting values,  max (min), of the CKM matrix
 elements for   model $(1)$: dot-dot-dashed line (dot-dash-dashed line) for $N_{c}^{eff}=0.66$. 
Solid line (dotted line) for $N_{c}^{eff}=2.69$.  Dashed line (dot-dashed line) for $N_{c}^{eff}=2.84$.}
\label{fig6}
\end{figure}

\begin{figure}
\centering\includegraphics[height=7cm,clip=true]{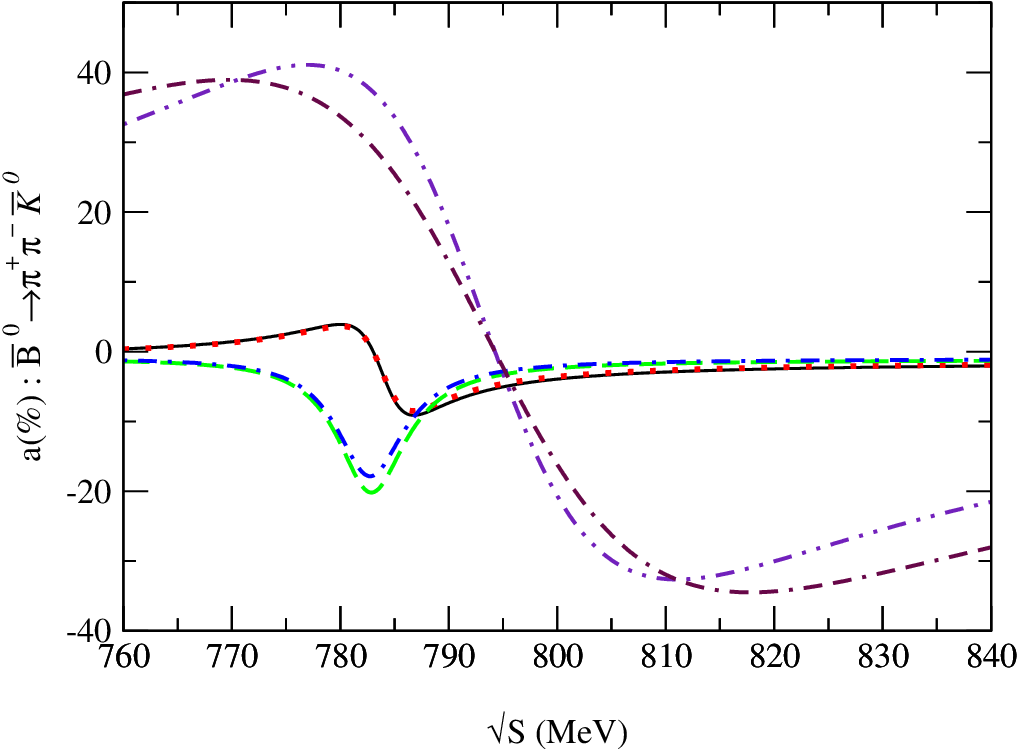}
\caption{ $CP$  violating asymmetry, $a$, for ${\bar B}^{0} \rightarrow \pi^{+} \pi^{-} {\bar K}^{0}$, 
for $k^{2}/m_{b}^{2}=0.5$, for $ N_{c}^{eff}= 0.61,2.65,2.82$  and for limiting values, max (min), of
 the  CKM matrix elements for  model $(1)$:  dot-dot-dashed line (dot-dash-dashed line) for $N_{c}^{eff}=0.61$. Solid 
line (dotted line) for $N_{c}^{eff}=2.65$. Dashed line (dot-dashed line) for $N_{c}^{eff}=2.82$.}
\label{fig7}
\end{figure}

\begin{figure}
\centering\includegraphics[height=7cm,clip=true]{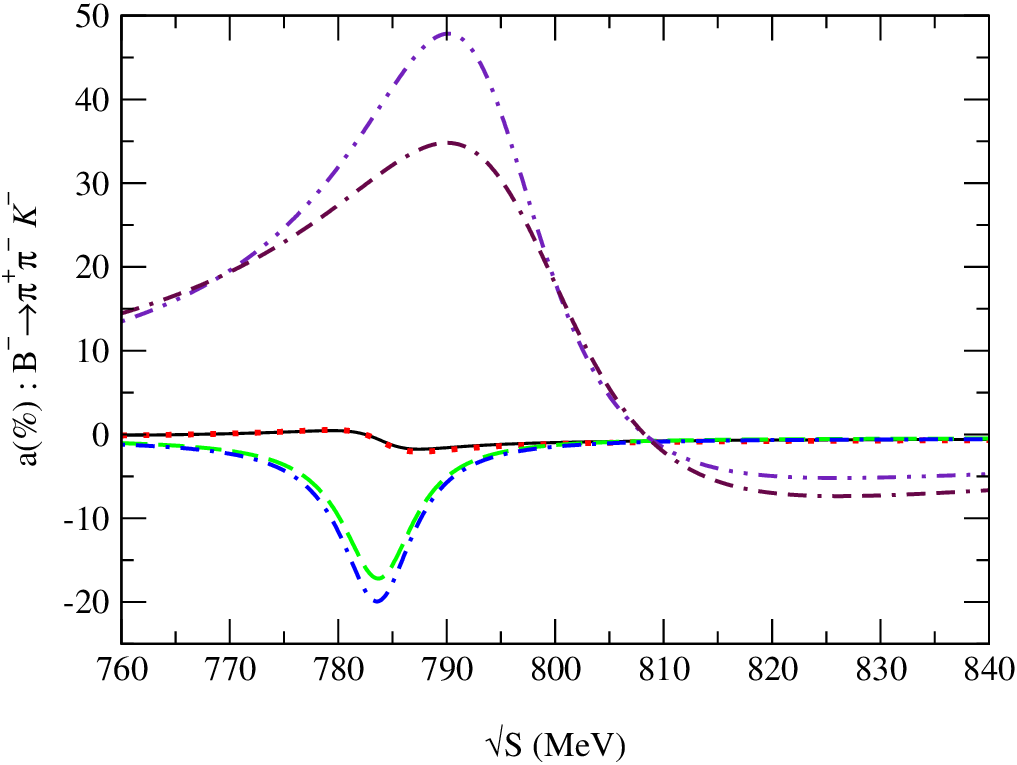}
\caption{ $CP$  violating asymmetry, $a$, for $B^{-} \rightarrow \pi^{+} \pi^{-}  K^{-}$, 
for $k^{2}/m_{b}^{2}=0.3$, for $ N_{c}^{eff}=0.66,2.69,2.84$  and for limiting values, max (min), of the CKM matrix
 elements for   model $(1)$:   dot-dot-dashed line (dot-dash-dashed line) for $N_{c}^{eff}=0.66$. 
Solid line (dotted line) for $N_{c}^{eff}=2.69$.  Dashed line (dot-dashed line) for $N_{c}^{eff}=2.84$.}
\label{fig8}
\end{figure}

\begin{figure}
\centering\includegraphics[height=7cm,clip=true]{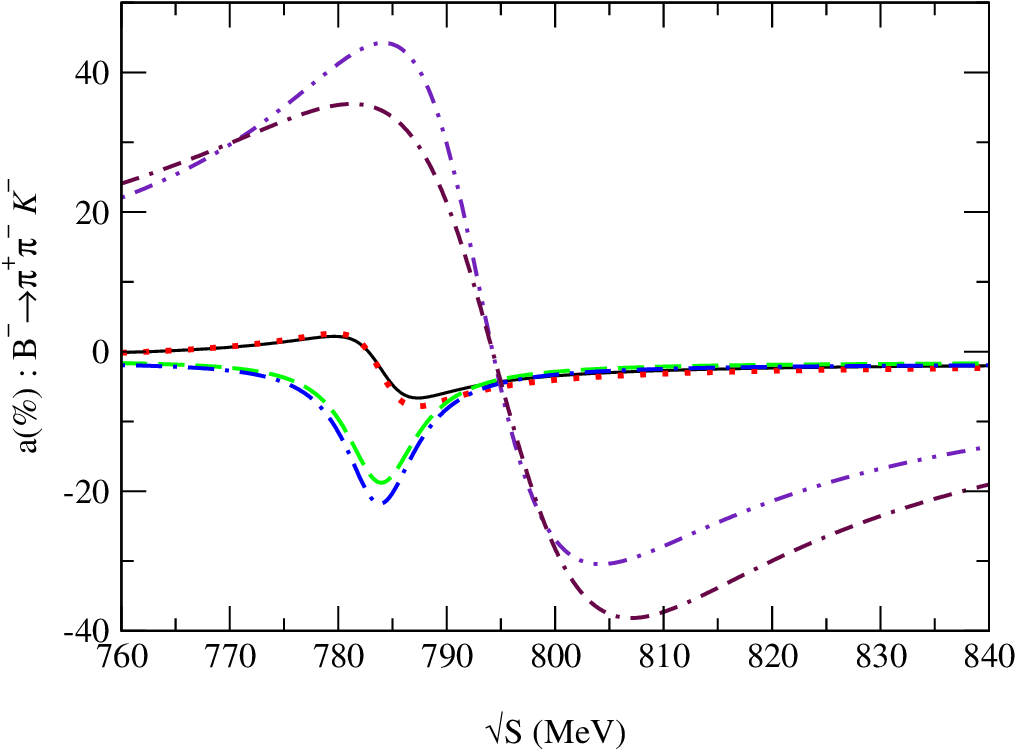}
\caption{ $CP$  violating asymmetry, $a$, for $B^{-} \rightarrow \pi^{+} \pi^{-} K^{-}$,
 for $k^{2}/m_{b}^{2}=0.5$, for $ N_{c}^{eff}=0.61,2.65,2.82$  and for limiting values, max (min), of
 the  CKM matrix elements for  model $(1)$:   dot-dot-dashed line (dot-dash-dashed line) for $N_{c}^{eff}=0.61$. Solid 
line (dotted line) for $N_{c}^{eff}=2.65$. Dashed line (dot-dashed line) for $N_{c}^{eff}=2.82$.}
\label{fig9}
\end{figure}

\begin{figure}
\centering\includegraphics[height=7cm,clip=true]{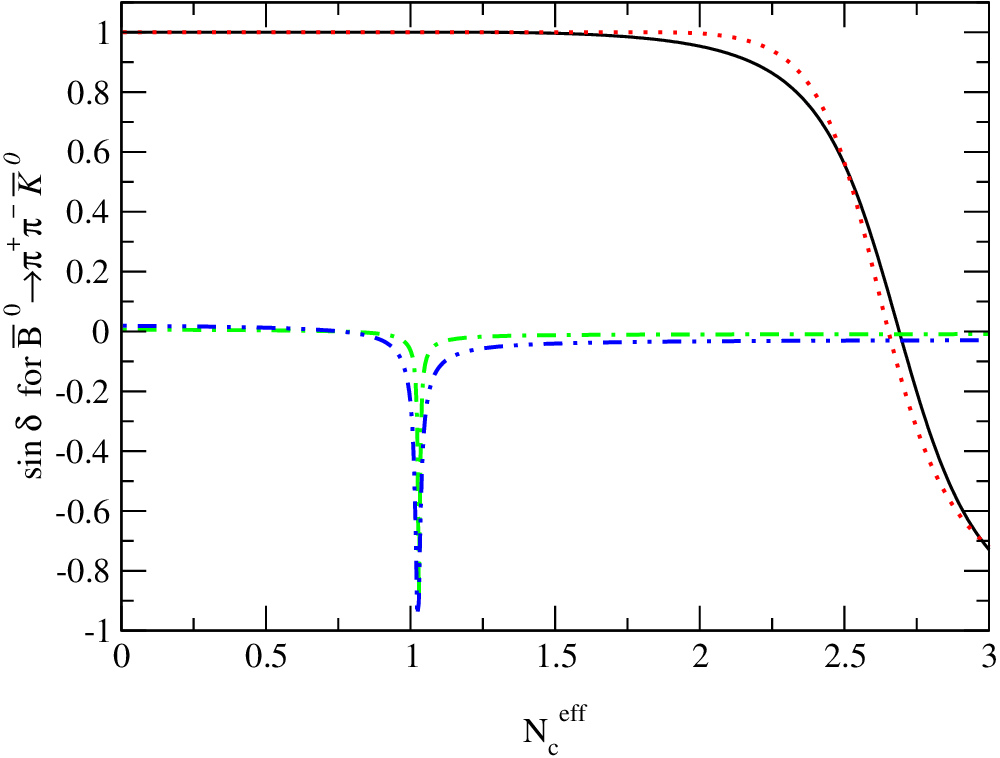}
\caption{$\sin\delta$,  as a function of $N_{c}^{eff}$, for ${\bar B}^{0} \rightarrow
 \pi^{+} \pi^{-} {\bar K}^{0}$, for $k^{2}/m_{B}^{2}=0.3(0.5)$ and for  model $(1)$.   The solid (dotted) line at
 $\sin \delta=+1$ corresponds to the case    $\tilde{\Pi}_{\rho \omega}=(-3500;-300)$, where     
 $\rho - \omega$ mixing is included. The  dot-dashed (dot-dot-dashed) line corresponds to 
 $\tilde{\Pi}_{\rho \omega}=(0;0)$, where   $\rho - \omega$ mixing is not included. }
\label{fig10}
\end{figure}

\begin{figure}
\centering\includegraphics[height=7cm,clip=true]{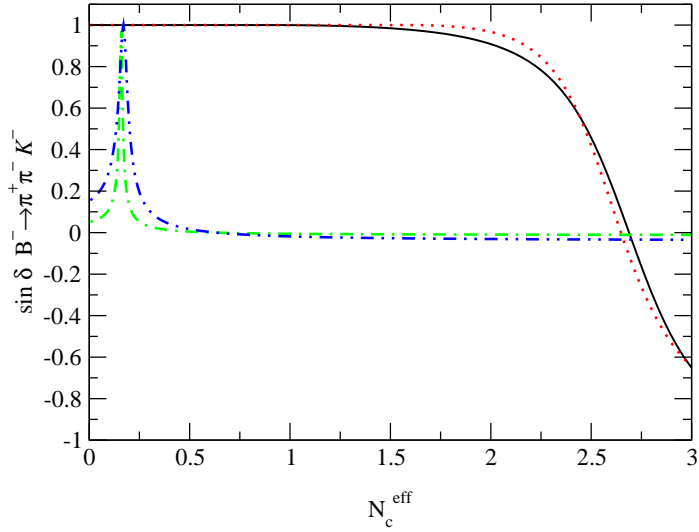}
\caption{$\sin\delta$,  as a function of $N_{c}^{eff}$ for $B^{-} \rightarrow
 \pi^{+} \pi^{-} K^{-}$, for $k^{2}/m_{B}^{2}=0.3(0.5)$ and for  model $(1)$.   The solid (dotted) line at
 $\sin \delta=+1$ corresponds to the case    $\tilde{\Pi}_{\rho \omega}=(-3500;-300)$, where     
 $\rho - \omega$ mixing is included. The  dot-dashed (dot-dot-dashed) line corresponds to 
 $\tilde{\Pi}_{\rho \omega}=(0;0)$, where   $\rho - \omega$ mixing is not included.}
\label{fig11}
\end{figure}

\begin{figure}
\centering\includegraphics[height=7cm,clip=true]{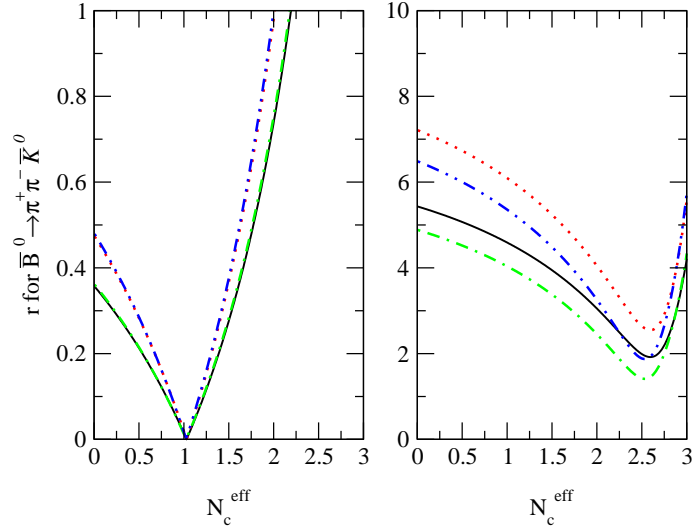}
\caption{The ratio of penguin to tree amplitudes, $r$, as a function of $N_{c}^{eff}$, for ${\bar B}^{0} \rightarrow
 \pi^{+} \pi^{-} {\bar K}^{0}$, for $k^{2}/m_{B}^{2}=0.3(0.5)$, 
for limiting values of the CKM matrix elements $(\rho,\eta)$ max(min),  for  $\tilde{\Pi}_{\rho
 \omega}=(-3500;-300)(0,0)$, (i.e. with(without) $\rho - \omega$  mixing) and for   model $(1)$. Figure 12a
 (left): for $\tilde{\Pi}_{\rho \omega}=(0;0)$,  solid line (dotted line) for $k^{2}/m_{B}^{2}=0.3$ and
 $(\rho,\eta)$ max(min). Dot-dashed line (dot-dot-dashed line) for  $k^{2}/m_{B}^{2}=0.5$ and $(\rho,\eta)$
 max(min). Figure 12b (right): same caption but for  $\tilde{\Pi}_{\rho \omega}=(-3500;-300)$. }
\label{fig12}
\end{figure}

\begin{figure}
\centering\includegraphics[height=7cm,clip=true]{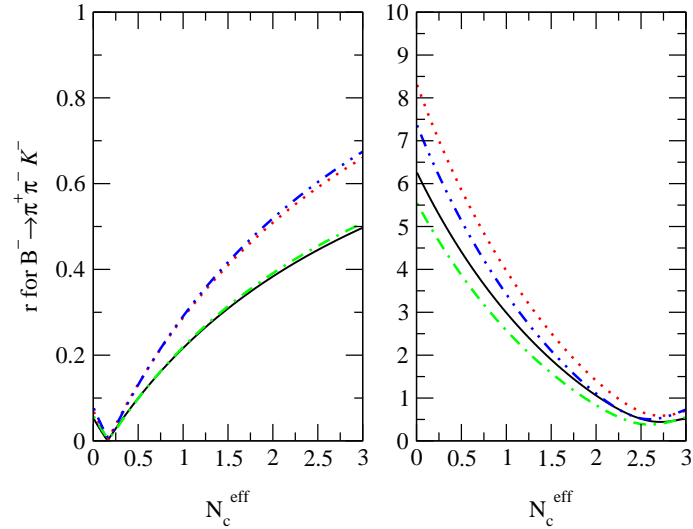}
\caption{The ratio of penguin to tree amplitudes, $r$, for $B^{-} \rightarrow
 \pi^{+} \pi^{-} K^{-}$.  Same caption for 
Figure 13a (left) and Figure 13b (right) as in  Fig.~\ref{fig12}.}
\label{fig13}
\end{figure}

%
With the factorized decay amplitudes, we can compute the decay rates by using   the following expression~\cite{ref25},
\begin{equation}\label{eq1.40}
\Gamma(B \rightarrow VP)=\frac{|\vec{p}_{\rho}|^{3}}{8\pi m_{V}^{2}}
\left|\frac{A(B \rightarrow VP)}{\epsilon_{V} \cdot p_{B}}\right|^{2}\ , 
\end{equation}
where $\vec{p}_{\rho}$ is the c.m. momentum of the decay particles  defined  as,
\begin{equation}\label{eq1.41}
|\vec{p}_{\rho}|=\frac{ \sqrt{ [m_{B}^{2}-(m_{1}+m_{2})^{2}][m_{B}^{2}-(m_{1}-m_{2})^{2}]}}{2m_{B}}\ .
\end{equation}
$m_{1} (m_{2})$ is the mass of the vector (pseudo-scalar) $V(P)$ particle, $\epsilon_{V}$ is the polarization 
vector and $A(B \rightarrow VP)$ is the decay amplitude given by, 
\begin{equation}\label{eq1.42}
A(B \rightarrow VP)=\frac{G_{F}}{\sqrt{2}} \sum_{i=1,10}V_{s}^{T,P}a_{i}\langle VP | O_{i}| B \rangle\ ,
\end{equation}
where the effective parameters, $a_{i}$, which are involved in the decay amplitude,  are the following
combinations of effective Wilson coefficients:
\begin{equation}\label{eq1.44}
a_{2j}=C_{2j}^{\prime}+\frac{1}{N_{c}^{eff}}C_{2j-1}^{\prime},   \;\;\; a_{2j-1}=C_{2j-1}^{\prime}+\frac{1}
{N_{c}^{eff}}C_{2j}^{\prime}, \;\;  {\rm for} \;\; j=1, \cdots, 5\ .
\end{equation}
All other variables in Eq.~(\ref{eq1.42}) have been introduced earlier.
In the Quark Model, the diagram (Fig.~\ref{fig5} top) gives the main contribution to 
 the $B  \rightarrow \rho^{0} K$ decay. In our case, to be consistent, we should also take into account the 
 $\rho - \omega$ 
mixing contribution (Fig.~\ref{fig5} bottom)  when we calculate the branching ratio, since we are working 
to the first order of  isospin
 violation. The application is straightforward and we obtain the branching ratio for $B \rightarrow \rho^{0} K$:
\begin{multline}\label{eq1.43}
{\rm BR}(B  \rightarrow \rho^{0} K)=\frac{G_{F}^{2}|\vec{p}_{\rho}|^{3}}{\alpha_k \pi \Gamma_{B}}\Bigg|
\bigg[V_{s}^{T}A^{T}_{\rho^{0}}(a_{1},a_{2})-V_{s}^{P}A^{P}_{\rho^{0}}(a_{3}, \cdots, a_{10})\bigg]    \\
+\bigg[V_{s}^{T}A^{T}_{\omega}(a_{1},a_{2})-V_{s}^{P}A^{P}_{\omega}(a_{3}, \cdots, a_{10})\bigg]\frac{
\tilde{\Pi}_{\rho \omega}}{(s_{\rho}-m_{\omega}^{2})+im_{\omega}\Gamma_{\omega}}\Bigg|^{2}\ .
\end{multline}
In Eq.~(\ref{eq1.43}) $G_{F}$ is the Fermi constant, $\Gamma_{B}$ is  the total width $B$ decay, and  $\alpha_k$ is 
  an integer related
to the given decay. $A^{T}_{V}$ and $A^{P}_{V}$ are the tree and penguin amplitudes which respect
quark interactions in the $B$ decay.     $V_{s}^{T,P}$ (in Eq.~(\ref{eq1.42})) or $V_{s}^{T},V_{s}^{P}$ 
(in Eq.~(\ref{eq1.43})) represent the CKM matrix elements involved in  the tree and penguin diagram, respectively: 
\begin{equation}
V_{s}^{T}  =|V_{ub}V_{us}^{\star}| \;\; {\rm for} \;\;\;\; i=1,2\ ,  \;\;\;\; {\rm and} \;\;\; 
V_{s}^{P}  =|V_{tb}V_{ts}^{\star}| \;\; {\rm for} \;\;\;\; i=3,\cdots,10\ .
\end{equation}
%
%
%
\subsection{Calculational details}\label{part5.2}
%
%
In this section, we enumerate the theoretical decay amplitudes.
We shall analyze five  $b$ into $s$ transitions. Two of them involve $\rho-\omega$ mixing. These are
$B^{-} \rightarrow \rho^{0} K^{-}$ and ${\Bar B}^{0} \rightarrow \rho^{0} {\Bar K}^{0}$. Two other decays are 
${\Bar B}^{0} \rightarrow \rho^{-} K^{+}$ and $B^{-} \rightarrow \rho^{-} {\Bar K}^{0}$ and the last one
 is $B^{-} \rightarrow \omega K^{-}$. We list  in the following, 
the tree and penguin amplitudes which appear in the given transitions. 
\newline
\noindent For  the decay $B^{-} \rightarrow \rho^{0} K^{-}$ ($\alpha_k = 32$ in Eq.~(\ref{eq1.43})),
\begin{multline}\label{eq1.45}
\sqrt{2}A^{T}_{\rho}(a_{1},a_{2})  = a_{1}f_{\rho}F_{1}(m_{\rho}^{2})+a_{2}f_{K}A_{0}(m_{K}^{2})\ , 
\hspace{16em}
\end{multline}
\begin{multline}\label{eq1.46}
\sqrt{2}A^{P}_{\rho}(a_{3},  \cdots,   a_{10})  =
f_{\rho}F_{1}(m_{\rho}^{2}) \biggl\{ \frac{3}{2}(a_{7}+a_{9}) \biggr\} \\
+f_{K} A_{0}  (m_{K}^{2}) \Biggl\{ a_{4} + a_{10} -2(a_{6}+a_{8}) \biggl[ \frac{m_{K}^{2} }
{(m_{u}+m_{s})(m_{b}+m_{u})}\biggr] \Biggl\}\ ;
\end{multline}
\noindent for  the decay $B^{-} \rightarrow \omega  K^{-}$ ($\alpha_k = 32$ in Eq.~(\ref{eq1.43})),
\begin{multline}\label{eq1.47}
\sqrt{2}A^{T}_{\omega}(a_{1},a_{2})  = a_{1}f_{\rho}F_{1}(m_{\rho}^{2})+a_{2}f_{K}A_{0}(m_{K}^{2})\ ,
 \hspace{16em}
\end{multline}
\begin{multline}\label{eq1.48}
\sqrt{2}A^{P}_{\omega}(a_{3},  \cdots,   a_{10})  = f_{\rho} F_{1}(m_{\rho}^{2})
\bigg\{ 2(a_{3}+a_{5})+\frac{1}{2}(a_{7}
+a_{9})  \bigg\} \\
 + f_{K}A_{0}(m_{K}^{2})\Biggl\{ -2 (a_{8}+a_{6}) \biggl[ \frac{m_{K}^{2}}{(m_{u}+m_{s})(m_{b}+m_{u})}\biggr] + 
 a_{4} + a_{10}  \Biggr\}\ ;  
\end{multline}
\noindent for  the decay  ${\Bar B}^{0} \rightarrow \rho^{0} {\Bar K}^{0}$ ($\alpha_k = 32$ in Eq.~(\ref{eq1.43})),
\begin{multline}\label{eq1.49}
\sqrt{2}A^{T}_{\rho}(a_{1},a_{2})  = a_{1}f_{\rho}F_{1}(m_{\rho}^{2})\ , 
\hspace{22em} 
\end{multline}
\begin{multline}\label{eq1.50}
\sqrt{2}A^{P}_{\rho}(a_{3},  \cdots,   a_{10})  = 
f_{\rho}F_{1}(m_{\rho}^{2}) \biggl\{  \frac{3}{2}(a_{7}+a_{9})\biggr\} \\
+f_{K} A_{0}(m_{K}^{2}) \Biggl\{ a_{4} - (2 a_{6} - a_{8}) \biggl[ \frac{m_{K}^{2}}
{(m_{s}+m_{d})(m_{b}+ m_{d})}\biggr] -\frac{1}{2}a_{10} \Bigg\}\ ;
\end{multline}
\noindent for  the decay  ${\Bar B}^{0} \rightarrow \omega {\Bar K}^{0}$ ($\alpha_k = 32$ in Eq.~(\ref{eq1.43})),
\begin{multline}\label{eq1.51}
\sqrt{2}A^{T}_{\omega}(a_{1},a_{2})  = a_{1} f_{\rho}F_{1}(m_{\rho}^{2})\ ,  
\hspace{22em} 
\end{multline}
\begin{multline}\label{eq1.52}
 \sqrt{2}A^{P}_{\omega}(a_{3},  \cdots,   a_{10})  = 
f_{\rho}F_{1}(m_{\rho}^{2}) \biggl\{ 2 (a_{3}+ a_{5})  + \frac{1}{2}(a_{7}+a_{9}) \biggr\} \\
+f_{K} A_{0}(m_{K}^{2}) \Biggl\{ a_{4} - (2 a_{6} - a_{8}) \biggl[ \frac{m_{K}^{2}}
{(m_{s}+m_{d})(m_{b}+ m_{d})}\biggr] -\frac{1}{2}a_{10} \Biggr\} \ ;  
\end{multline}
\noindent  for  the decay   $B^{-} \rightarrow \rho^{-} {\Bar K}^{0}$ ($\alpha_k = 16$ in Eq.~(\ref{eq1.43})),
\begin{multline}\label{eq1.53}
A^{T}_{\rho}(a_{1},a_{2})  = a_{2}f_{\rho}F_{1}(m_{\rho}^{2})\ , 
\hspace{23em}
\end{multline}
\begin{multline}\label{eq1.54}
A^{P}_{\rho}(a_{3},\cdots, a_{10})  = 
 f_{K} A_{0}(m_{K}^{2}) \Biggl\{ a_{4} -\frac{1}{2}a_{10}-(2 a_{6}
 - a_{8}) \biggl[ \frac{m_{K}^{2}}{(m_{s}+m_{d})(m_{b}+ m_{d})}\biggr]  \Biggr\}\ ; 
\end{multline}
\noindent for  the decay  ${\Bar B}^{0} \rightarrow \rho^{+} K^{-}$ ($\alpha_k = 16$ in Eq.~(\ref{eq1.43})),
\begin{multline}\label{eq1.55}
A^{T}_{\rho}(a_{1},a_{2})  = a_{2} f_{K} A_{0}(m_{K}^{2})\ , 
\hspace{23em}
\end{multline}
\begin{multline}\label{eq1.56}
A^{P}_{\rho}(a_{3},\cdots, a_{10})  = 
f_{K} A_{0}(m_{K}^{2}) \Biggl\{a_{4} + a_{10}  - 2( a_{6} + a_{8}) \biggl[ \frac{m_{K}^{2}}
{(m_{s}+m_{u})(m_{b}+ m_{u})}\biggr] \Biggr\}\ . 
\end{multline}
Moreover, we can calculate the ratio between two branching ratios, in which the uncertainty caused by 
many systematic errors is removed.  We define the ratio $R$  as: 
\begin{eqnarray}
R= \frac{{\rm BR}(B^{0} \rightarrow \rho^{\pm} K^{\mp})}{{\rm BR}(B^{\pm} \rightarrow \rho^{0} K^{\pm})}\ ,
\end{eqnarray}
and, without taking  into account the penguin contribution, one has,
\begin{eqnarray}
R=\frac{2 \Gamma_{B^{+}}}{ \Gamma_{B^{0}}} \bigg| \bigg( 1+  \frac{a_{1} f_{\rho}F_{1}(m_{\rho}^{2})}{ a_{2} 
f_{K} A_{0}(m_{K}^{2})
} \bigg) \bigg(1+\frac{\tilde{\Pi}_{\rho \omega}}{(s_{\rho}-m_{\omega}^{2})+im_{\omega}
\Gamma_{\omega}}\bigg) \bigg|^{-2}\ .
\end{eqnarray}
%
%
\subsection{Numerical results}\label{part5.3}
%

The numerical values for  the CKM matrix elements  $V_{s}^{T,P}$, 
the $\rho-\omega$ mixing amplitude  $\tilde{\Pi}_{\rho \omega}$, and  the 
 particle masses  $m_{V,P}$, which appear  in Eq.~(\ref{eq1.43}),  
have been all   reported in Section~\ref{part3}. The Fermi constant
is taken  to  be $G_{F}= 1.166391 \times 10^{-5} \;  {\rm GeV}^{-2}$~\cite{ref16}, and for the total width $B$ decay, 
$\Gamma_{B}(= 1/\tau_{B})$, we use 
 the  world average $B$ lifetime values (combined results from ALEPH, Collider Detector at Fermilab (CDF), DELPHI, L3, OPAL and SLAC Large Detector (SLD))~\cite{ref26}:
\begin{align}\label{eq1.57}
\tau_{B^{0}} & = 1.546 \pm 0.021 \;{\rm ps}\ , \nonumber \\
\tau_{B^{+}} & = 1.647 \pm 0.021 \;{\rm ps}\ .
\end{align}

To compare the 
theoretical results with experimental data, as well as  to  determine the constraints on the effective number
of color, $N_{c}^{eff}$, the form factors, and the CKM matrix parameters, we shall apply the experimental 
branching ratios
collected at CLEO~\cite{ref29}, BELLE~\cite{ref30,ref31,ref32} and BABAR~\cite{ref33,ref34} factories. All the 
experimental values are summarized  in Table~\ref{tab6}.

\begin{figure}
\centering\includegraphics[height=7cm,clip=true]{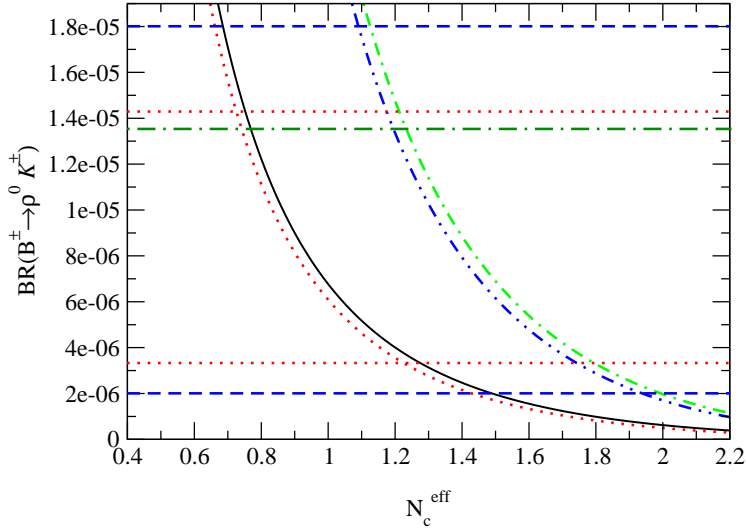}
\caption{Branching ratio for $B^{\pm} \rightarrow \rho^{0}  K^{\pm}$, for models  $1(2)$, $k^{2}/m_{B}^{2}=0.3$ 
and limiting values of the  CKM matrix elements. Solid line (dotted line) for  model $(1)$ and max (min) CKM matrix 
elements. Dot-dashed line (dot-dot-dashed line) for   model $(2)$ and max (min) CKM matrix elements. Notation: 
horizontal 
dotted  line: CLEO data; dashed line: BABAR data; dot-dashed line: BELLE data.}
\label{fig14}
\end{figure}

\begin{figure}
\centering\includegraphics[height=7cm,clip=true]{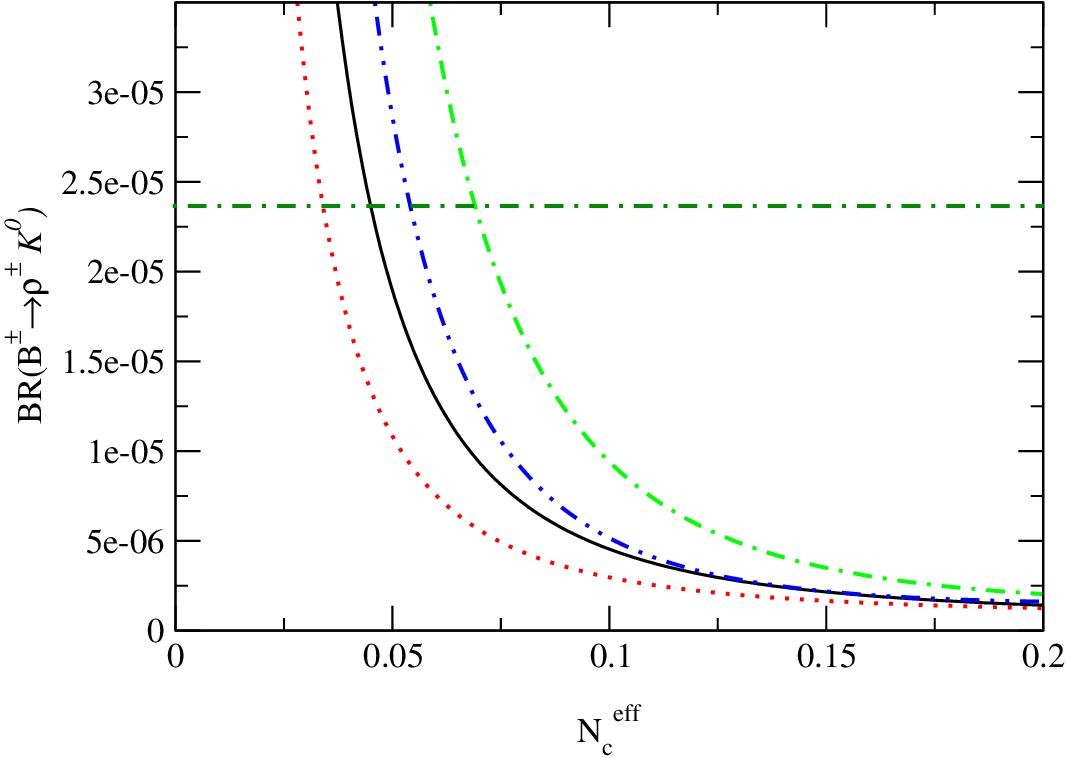}
\caption{Branching ratio for $B^{\pm} \rightarrow \rho^{\pm}K^{0}$, for models  $1(2)$, $k^{2}/m_{B}^{2}=0.3$
 and limiting values of the  CKM matrix elements. Solid line (dotted  line) for  model $(1)$ and max (min) CKM 
matrix elements. Dot-dashed line (dot-dot-dashed line) for  model $(2)$ and max (min) CKM matrix elements. Same 
notation as in Fig.~\ref{fig14}, but only experimental upper limits are available.}
\label{fig15}
\end{figure}

\begin{figure}
\centering\includegraphics[height=7cm,clip=true]{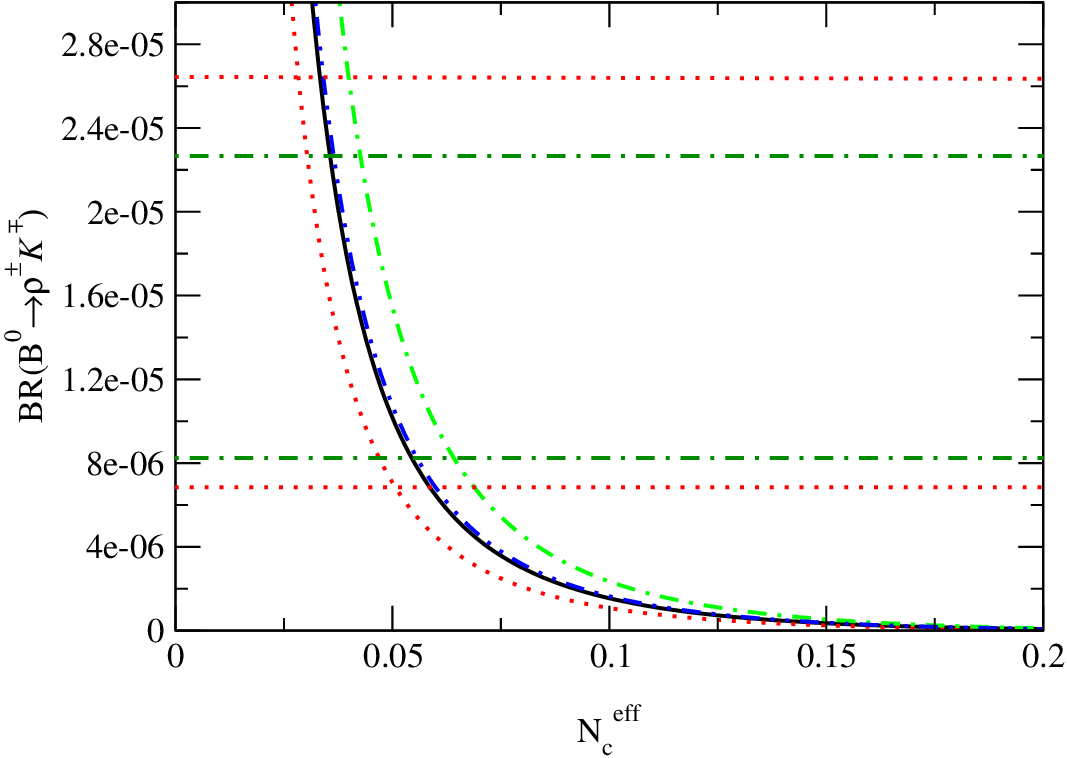}
\caption{Branching ratio for $B^{0} \rightarrow \rho^{\pm}K^{\mp}$, for models  $1(2)$, $k^{2}/m_{B}^{2}=0.3$
 and limiting values of the  CKM matrix elements. Solid line (dotted line) for  model $(1)$ and max (min) CKM 
matrix elements. Dot-dashed line (dot-dot-dashed line) for  model $(2)$ and max (min) CKM matrix elements.
Same notation as in Fig.~\ref{fig14}.}
\label{fig16}
\end{figure}

\begin{figure}
\centering\includegraphics[height=7cm,clip=true]{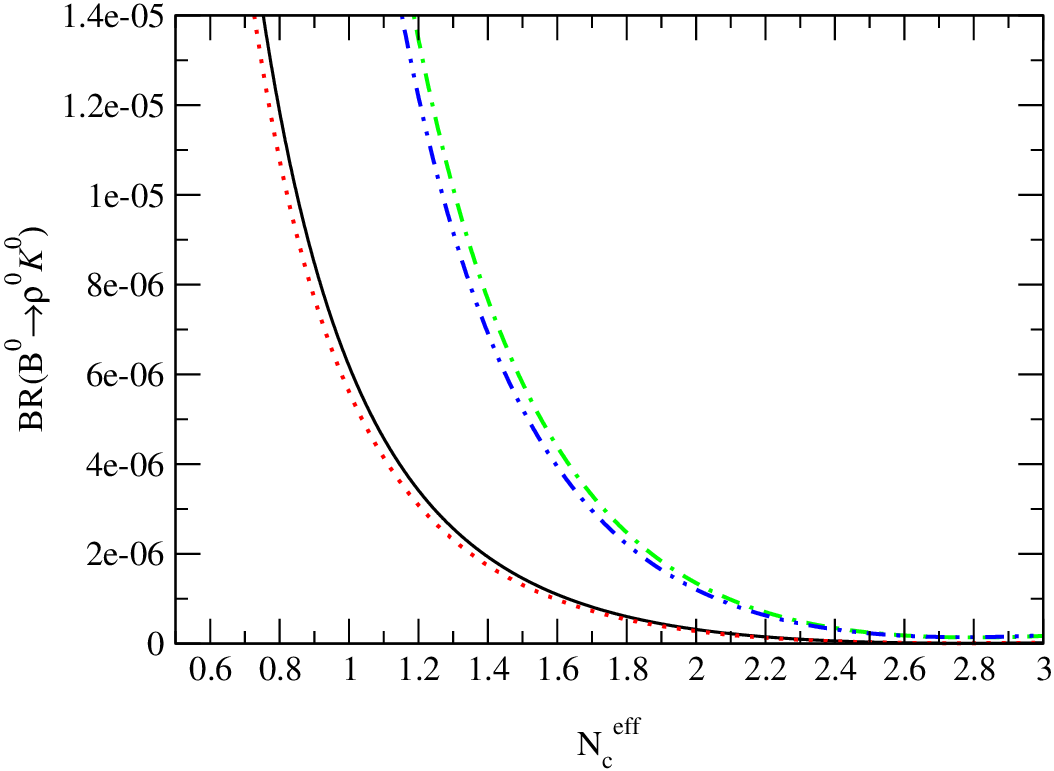}
\caption{Branching ratio for $B^{0} \rightarrow \rho^{0}K^{0}$, for models  $1(2)$, $k^{2}/m_{B}^{2}=0.3$
 and limiting values of the  CKM matrix elements. Solid line (dotted line) for  model $(1)$ and max (min) CKM 
matrix elements. Dot-dashed line (dot-dot-dashed line) for  model $(2)$ and max (min) CKM matrix elements.}
\label{fig17}
\end{figure}

\begin{figure}
\centering\includegraphics[height=7cm,clip=true]{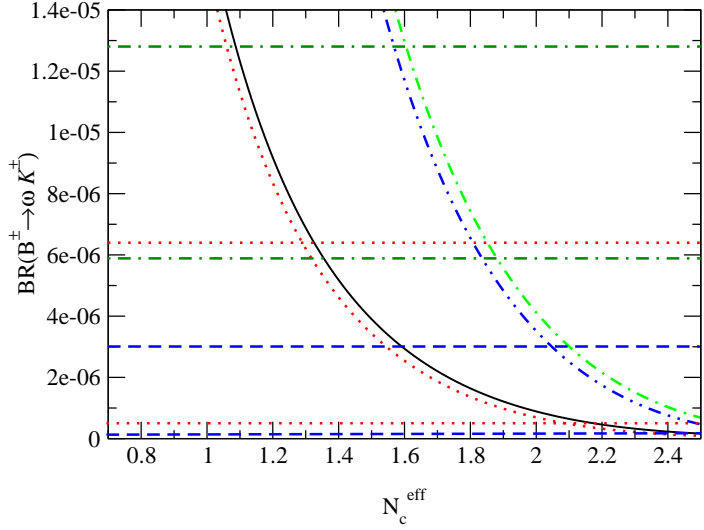}
\caption{Branching ratio for $B^{\pm} \rightarrow \omega K^{\pm}$, for models  $1(2)$, $k^{2}/m_{B}^{2}=0.3$
 and limiting values of the  CKM matrix elements. Solid line (dotted  line) for  model $(1)$ and max (min) CKM 
matrix elements. Dot-dashed line (dot-dot-dashed line) for  model $(2)$ and max (min) CKM matrix elements. Same 
notation as in Fig.~\ref{fig14}.}
\label{fig18}
\end{figure}

In order to determine the  range of $N_{c}^{eff}$  available for calculating the  $CP$  violating parameter, $a$, in 
$B^{\pm,0} \rightarrow \rho^{0} K^{\pm,0}$, we have calculated the branching ratios for $B^{\pm} \rightarrow
\rho^{0} K^{\pm}$, $B^{\pm} \rightarrow \rho^{\pm} K^{0}$, $B^{0} \rightarrow \rho^{\pm} K^{\mp}$,
$B^{0} \rightarrow \rho^{0} K^{0}$, and  $B^{\pm} \rightarrow \omega K^{\pm}$. 
We show all  the 
results in Figs.~\ref{fig14},~\ref{fig15},~\ref{fig16},~\ref{fig17}, and~\ref{fig18}, where  branching ratios are plotted as a function of $N_{c}^{eff}$ for models $(1)$ and $(2)$ 
(different form factors are used in  
 models $(1)$ and $(2)$). By taking experimental data from CLEO, BABAR and BELLE 
Collaborations, listed in Table~\ref{tab6}, and comparing theoretical predictions with experimental results, we expect
to extract the allowed range of $N_{c}^{eff}$ in $B \rightarrow \rho K$ and to make the dependence on the 
 form factors
explicit between the two classes of models: models $(1,3)$ and $(5)$, and models $(2)$ and $(4)$.
We shall mainly use the CLEO data, since the BABAR and BELLE data are (as yet) 
less numerous and accurate. An exception will be made for the branching ratio $B^{\pm} \rightarrow \omega K^{\pm}$, 
where
we shall take  the BELLE data for our analysis since they are the most accurate and most recent 
measurements in that case.
Nevertheless, we shall
also apply all of them to check the agreement between all  the branching ratio data. The CLEO, BABAR and BELLE 
Collaborations give almost the same experimental branching ratios for 
all the investigated decays except for the decay $B^{-} \rightarrow \omega K^{-}$. In this later case,  we observe
a strong 
disagreement between all of them since they provide experimental data in a range  from $0.1 \times  10^{-6}$ to
$12.8 \times 10^{-6}$. 
Finally, it is evident that numerical results are very sensitive to uncertainties coming from the experimental data
and from  the factorization approach applied to calculate hadronic matrix
elements in the $B \rightarrow K$ transition.  Moreover, for  $B \rightarrow \rho K$,
the data are less numerous than for $B \rightarrow \rho \pi$, so we cannot expect to get a very accurate range 
of $N_{c}^{eff}$.

For the branching ratio $B^{\pm} \rightarrow \rho^{0} K^{\pm}$ (Fig.~\ref{fig14}) we found a large range of values
of $N_{c}^{eff}$ and CKM matrix elements over which the theoretical results are consistent with experimental data from
CLEO, BABAR and BELLE. Each of 
 the models, $(1,2,3,4)$ and $(5)$, gives an allowed range of $N_{c}^{eff}$. Even though 
strong differences appear  between the two classes of models, because of the different used form factors, we are 
not able to draw strong conclusions about the dependence on the 
 form factors. For the branching ratio $B^{\pm} \rightarrow \rho^{\pm}
K^{0}$, (Fig.~\ref{fig15}),  BELLE gives only an upper branching ratio limit whereas BABAR and CLEO do  not. Our
predictions are still consistent  with  the experimental data for all models, for a large range of $N_{c}^{eff}$. In
 this
case, the numerical results for models $(1)$ and $(2)$ are very close to each other and, we 
need new data to constrain our calculations.

If we consider our results for the branching ratio $B^{0} \rightarrow \rho^{\pm} K^{\mp}$ (plotted in Fig.~\ref
{fig16}), there is  agreement  between the 
experimental results from CLEO and BELLE (no data from BABAR)  and our theoretical predictions
at very low values   of $N_{c}^{eff}$ and the CKM matrix elements. All the models $(1,2,3,4)$ and $(5)$, give  
branching values within the range of branching ratio measurements if $N_{c}^{eff}$ is less than $0.07$. The tiny 
difference observed between models $(1)$ and $(2)$ comes from the form factor $A_{0}(k^{2})$  (where
$A_{0}(k^{2})$ refers to the $B$ to $\rho$ transition taken at $k^{2}=m_{K}^{2}$) since in that  case,
the amplitude computed involves only the form factor  $A_{0}(k^{2})$.
For  the branching
ratio $B^{0} \rightarrow \rho^{0} K^{0}$ shown in Fig.~\ref{fig17}, neither CLEO, BABAR nor BELLE give experimental
results. Nevertheless, from models $(1)$ and $(2)$, it appears that this branching ratio is very sensitive to the 
magnitude of the form factor $F_{1}(k^{2})$ (in our case, $F_{1}(k^{2})$ is uncertain because  
 $h_{1}=0.360$ or $0.762$ in
models $(1)$ and $(2)$, respectively) since the tree contribution is only proportional to $F_{1}$. Moreover, from the 
range of allowed values of $N_{c}^{eff}$, we can estimate the upper limit of this branching ratio to be of the order
  $20 \times 10^{-6}$.
Finally, we focus
on the branching ratio $B^{\pm} \rightarrow \omega K^{\pm}$ which is plotted in Fig.~\ref{fig18} for models $(1)$ and 
$(2)$. We find that both the experimental and theoretical results are in agreement for a large range of values of 
$N_{c}^{eff}$. But, the models $(1)$ and $(2)$ do not give similar results because  the form factor $F_{1}$, applied
in these models, is very
different in both cases. Moreover, the dependence of the branching ratio on the CKM parameters $\rho$ and $\eta$ 
indicates that it would be possible to strongly constrain $\rho$ and $\eta$  with a very accurate experimental 
measurement for the  decay  $B^{-} \rightarrow \omega K^{-}$.

\begin{figure}
\centering\includegraphics[height=7cm,clip=true]{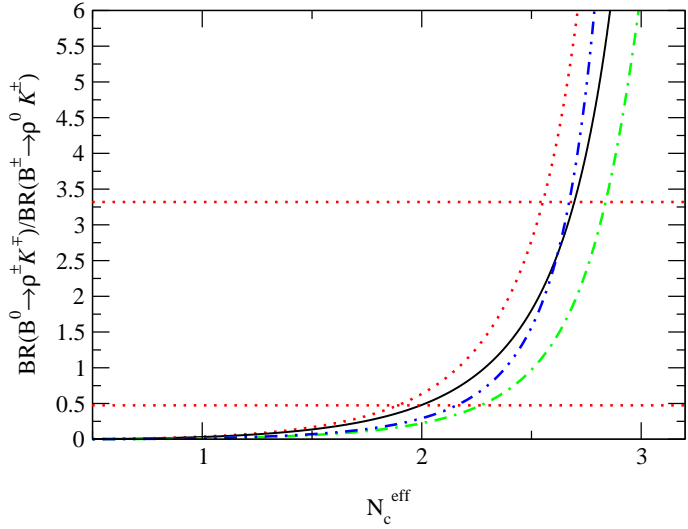}
\caption{The ratio of two $\rho K$ branching ratios versus $N_{c}^{eff}$  for models $1(2)$ 
and for limiting values of the CKM matrix elements: solid  line (dotted line) for  model $(1)$ with max (min) CKM
 matrix elements.  Dot-dashed line (dot-dot-dashed line) for   model $(2)$ with max (min) CKM matrix elements.  Same
 notation as in Fig.~\ref{fig14}.}
\label{fig19}
\end{figure}

To remove systematic errors in branching ratios given by the $B$ factories, we look at the ratio, $R$, between the two
following branching ratios: ${\rm BR}(B^{0} \rightarrow \rho^{\pm} K^{\mp})$ and ${\rm BR}(B^{\pm} \rightarrow 
\rho^{0} K^{\mp})$. The ratio is plotted in Fig.~\ref{fig19} as a function of $N_{c}^{eff}$, for models $(1)$ 
and $(2)$ and for limiting values of the CKM matrix elements. These results indicate that the ratio is very sensitive 
to both $N_{c}^{eff}$ and   to the magnitude of the form factors. The sensitivity increases with the value of
$N_{c}^{eff}$ and gives a large difference between models $(1,3)$ and $(5)$ and models $(2)$ and $(4)$. We found
that for a definite range of $N_{c}^{eff}$, all models investigated give a ratio consistent with the experimental
data from CLEO. It should be noted that $R$ is not very sensitive to the CKM matrix elements. Indeed, if  we only take
into account the tree contributions,  $R$ is independent of the CKM parameters $\rho$ and
$\eta$. The difference which appears comes from the penguin contribution and has to be taken into account in any 
approach since they are not negligible.

We have summarized for each model, each branching ratio and each set of limiting values of CKM matrix elements, 
the allowed range of $N_{c}^{eff}$ within which the 
 experimental data and numerical results are consistent. 
To determine the best
range of $N_{c}^{eff}$, we have to find some intersection of  values of $N_{c}^{eff}$ for each model and each 
set of CKM matrix elements, for which the theoretical and experimental results are consistent. Since the experimental
results are not numerous and not as accurate as one would like,
 it is more reasonable to fix the upper and lower limits of
$N_{c}^{eff}$ which allow us 
the maximum of agreement between the theoretical and experimental approaches. By using
the limiting values of the CKM matrix elements we show  in Table~\ref{tab15}, the range of 
allowed values of $N_{c}^{eff}$ with  $\rho-\omega$ mixing. Even though in our previous study for $B \rightarrow 
\rho \pi$, we have restricted ourselves to models $(2)$ and $(4)$ rather than models $(1,3)$ and $(5)$, here, we cannot
exclude one of the models $(1,2,3,4)$ and $(5)$ due to the lack of accurate experimental data.
We find that $N_{c}^{eff}$ should be in the following range: $0.66(0.61)< N_{c}^{eff} < 2.84(2.82)$, 
where the values outside and inside brackets correspond to the choice $k^{2}/m_{b}^{2}=0.3(0.5)$. Finally, if 
we take into account the allowed range of $N_{c}^{eff}$ determined for decays such as $B \rightarrow \rho \pi$ and 
$B \rightarrow \rho K$ we find a minimum global allowed range of $N_{c}^{eff}$ which should be in the range
$1.17(1.12)< N_{c}^{eff} < 1.63(1.77)$. 
%
%
\section{Summary and discussion}\label{part6}
%
%
We have studied direct $CP$ violation in decay process such as $B^{\pm,0} \rightarrow \rho^{0} K^{\pm,0} \rightarrow 
\pi^{+} \pi^{-} K^{\pm,0}$ with the inclusion of $\rho-\omega$ mixing. When the invariant mass of the $\pi^{+}\pi^{-}$
pair is in the vicinity of the $\omega$ resonance, it is found that the  $CP$  violating asymmetry, $a$, has a maximum
$a_{max}$. We have also investigated the branching ratios $B^{0} \rightarrow \rho^{0} K^{0}$,  $B^{0} \rightarrow 
\rho^{\pm} K^{\mp}$, $B^{\pm} \rightarrow \rho^{\pm} K^{0}$, $B^{\pm} \rightarrow \rho^{0} K^{\pm}$, and $B^{\pm}
 \rightarrow \omega K^{\pm}$. From our theoretical
results, we make comparisons with experimental data from the CLEO, BABAR and BELLE Collaborations. We have applied
five phenomenological models in order to show their  dependence on form factors, CKM matrix elements and the effective
parameter $N_{c}^{eff}$ in our approach.

To calculate the  $CP$  violating asymmetry, $a$, and the branching ratios, we started from the weak Hamiltonian in which 
the OPE separates hard and soft physical regimes. We  worked in the factorization approximation where 
the hadronic matrix elements are treated in some phenomenological quark models. The effective parameter, 
$N_{c}^{eff}$, was used in order to take into account, as well as possible, the non-factorizable effects
involved in $B \rightarrow \rho K$ decays. Although one must have some doubts about factorization, it has been
pointed out that it may be quite reliable in energetic weak decays~\cite{ref36}.

With the present work, we have explicitly shown that the direct  $CP$  
violating asymmetry is very sensitive to the CKM matrix 
elements, the magnitude of the form factors $A_{0}(k^{2})$ and $F_{1}(k^{2})$, and also to the effective parameter 
$N_{c}^{eff}$ (in order of increasing dependence). We have determined a range for the maximum asymmetry, $a_{max}$,
as a function of the parameter $N_{c}^{eff}$, the limits of CKM matrix elements and the choice of $k^{2}/m_{b}^{2}= 
0.3(0.5)$. For the decay ${\Bar B}^{0} \rightarrow \pi^{+} \pi^{-} {\Bar K}^{0}$ and from all models investigated, 
we found that the largest 
 $CP$  violating asymmetry varies from $+37\%(+55\%)$ to  $-20\%(-24\%)$. As regards $B^{-} \rightarrow \pi^{+}
\pi^{-} K^{-}$, one gets $+49\%(+46\%)$ to $-22\%(-25\%)$. For $B^{\pm,0} \rightarrow \pi^{+}\pi^{-} K^{\pm,0}$,
the sign of $a_{max}$ stays positive as long as the value of $N_{c}$ is less than $2.7$. 
In both decays, the ratio between asymmetry values which are 
taken at upper and lower limiting $\rho$ and $\eta$ values is mainly governed by the term $1/\sin \beta$. 
It appears also that the direct  $CP$  violating asymmetry is very sensitive to the form factors  at high 
values of $N_{c}^{eff}$.
 We underline
that without the inclusion of $\rho-\omega$ mixing, we would not have a large  $CP$  violating asymmetry, $a$, since $a$
 is 
proportional to both $\sin \delta$ and $r$. We found a critical point for which $\sin \delta$ reaches
the value $+1$, but at the same time, $r$ becomes very tiny. We emphazise that the
 advantage of $\rho-\omega$ mixing is the 
large strong phase difference  which varies extremely rapidly 
 near the $\omega$ resonance. In our calculations, we found that 
for  $B^{\pm,0} \rightarrow \pi^{+}\pi^{-} K^{\pm,0}$, the sign of $\sin \delta$ is positive until $N_{c}^{eff}$
reaches $2.69(2.65)$  when $k^{2}/m_{b}^{2}=0.3(0.5)$. Then, by measuring  $a$ for values of 
$N_{c}^{eff}$ lower than the limits given above, we can remove the phase uncertainty mod$(\pi)$ in the determination
of the CKM angle $\gamma$.

As regards theoretical results for the branching ratios $B^{\pm} \rightarrow \rho^{0} K^{\pm}$, $B^{\pm} \rightarrow
\rho^{\pm} K^{0}$, $B^{0} \rightarrow \rho^{\pm} K^{\mp}$, $B^{0} \rightarrow \rho^{0} K^{0}$ and  
$B^{\pm} \rightarrow \omega K^{\pm}$, we made comparison
with data from the 
CLEO (mainly), BABAR and BELLE (for $B^{\pm} \rightarrow \omega K^{\pm}$) Collaborations. We found that it is
 possible to have  agreement between the theoretical results and experimental branching ratio 
data for $B^{\pm} \rightarrow \rho^{0} K^{\pm}$, $B^{\pm} \rightarrow
\rho^{\pm} K^{0}$,  $B^{\pm} \rightarrow \omega K^{\pm}$, $B^{0} \rightarrow \rho^{\pm} K^{\mp}$, and $R$. 
For $B^{0} \rightarrow\rho^{0} K^{0}$, the lack of results does not allow us to draw 
conclusions. Only an estimation for the upper limit ($20 \times 10^{-6}$) has been determined. 
Nevertheless, we have determined a range of 
value of $N_{c}^{eff}$, $0.66(0.61)< N_{c}^{eff} < 2.84(2.82)$, inside of which the experimental data and theoretical 
calculations are consistent. We have to 
keep in mind that, because of the difficulty in  dealing  with non-factorizable effects
associated with final state interactions (FSI), which are more complex for decays involving an $s$ quark, 
we have weakly constrained the range of value of $N_{c}^{eff}$.

>From the  $CP$  violating asymmetry and the branching ratios, we expect to determine the CKM matrix elements. In order
to reach our aim, all uncertainties in our calculations have to be decreased: the transition form factors for $B
\rightarrow \rho$ and  $B \rightarrow K$ have to be well determined  and non-factorizable effects have to be treated
in the future by using generalized QCD factorization. Moreover, we strongly need more numerous and accurate 
experimental data in $B \rightarrow \rho K$ decays if we want to understand  direct  $CP$  violation in
$B$ decays better.
%
\subsubsection*{Acknowledgments}
%
This work was supported  in part by the Australian Research Council and the University of Adelaide.
\newpage
%

%
\newpage
%
%
\section*{Figure captions} 
%
%
\begin{itemize}
\item{ Fig.~\ref{fig1} Tree diagram, for $B$ decays.}
\item{ Fig.~\ref{fig2}  QCD-penguin diagram, for $B$ decays.}
\item{ Fig.~\ref{fig3} Electroweak-penguin diagram, for $B$ decays.}
\item{ Fig.~\ref{fig4} Electroweak-penguin diagram (coupling between $Z,\gamma$ and $W$), for $B$ decays.}
\item{ Fig.~\ref{fig5} $B$ decays without (upper) and with (lower) $\rho - \omega$ mixing.}
\item{ Fig.~\ref{fig6}   $CP$  violating asymmetry, $a$, for ${\bar B}^{0} \rightarrow \pi^{+} \pi^{-}  {\bar K}^{0}$,
for $k^{2}/m_{b}^{2}=0.3$, for $N_{c}^{eff}=0.66,2.69,2.84$  and for limiting values,  max (min), of the CKM matrix
 elements for   model $(1)$: dot-dot-dashed line (dot-dash-dashed line) for $N_{c}^{eff}=0.66$. 
Solid line (dotted line) for $N_{c}^{eff}=2.69$.  Dashed line (dot-dashed line) for $N_{c}^{eff}=2.84$.}
\item{ Fig.~\ref{fig7}   $CP$  violating asymmetry, $a$, for ${\bar B}^{0} \rightarrow \pi^{+} \pi^{-} {\bar K}^{0}$, 
for $k^{2}/m_{b}^{2}=0.5$, for $ N_{c}^{eff}=0.61,2.65,2.82$  and for limiting values, max (min), of
 the  CKM matrix elements for  model $(1)$:  dot-dot-dashed line (dot-dash-dashed line) for $N_{c}^{eff}=0.61$. Solid 
line (dotted line) for $N_{c}^{eff}=2.65$. Dashed line (dot-dashed line) for $N_{c}^{eff}=2.82$. }
\item{ Fig.~\ref{fig8}   $CP$  violating asymmetry, $a$, for $B^{-} \rightarrow \pi^{+} \pi^{-} K^{-}$, 
for $k^{2}/m_{b}^{2}=0.3$, for $ N_{c}^{eff}=0.66,2.69,2.84$  and for limiting values, max (min), of the CKM matrix
 elements for   model $(1)$:  dot-dot-dashed line (dot-dash-dashed line) for $N_{c}^{eff}=0.66$. 
Solid line (dotted line) 
for $N_{c}^{eff}=2.69$.  Dashed line (dot-dashed line) for $N_{c}^{eff}=2.84$.}
\item{ Fig.~\ref{fig9}   $CP$  violating asymmetry, $a$, for $B^{-} \rightarrow \pi^{+} \pi^{-} K^{-}$,
 for $k^{2}/m_{b}^{2}=0.5$, for $ N_{c}^{eff}=0.61,2.65,2.82$  and for limiting values, max (min), of
 the  CKM matrix elements for  model $(1)$:   dot-dot-dashed line (dot-dash-dashed line) for $N_{c}^{eff}=0.61$. Solid 
line (dotted line) for $N_{c}^{eff}=2.65$. Dashed line (dot-dashed line) for $N_{c}^{eff}=2.82$.}
\item{ Fig.~\ref{fig10}   $\sin\delta$, as a function of $N_{c}^{eff}$, for ${\Bar B}^{0} \rightarrow
 \pi^{+} \pi^{-} {\Bar K}^{0}$, for $k^{2}/m_{B}^{2}=0.3(0.5)$ and for  model $(1)$.   The solid (dotted) line at
 $\sin \delta=+1$ corresponds the case    $\tilde{\Pi}_{\rho \omega}=(-3500;-300)$, where     
 $\rho - \omega$ mixing is included. The  dot-dashed (dot-dot-dashed) line corresponds to 
 $\tilde{\Pi}_{\rho \omega}=(0;0)$, where   $\rho - \omega$ mixing is not included. }
\item{ Fig.~\ref{fig11}  $\sin\delta$,  as a function of $N_{c}^{eff}$, for $B^{-} \rightarrow
 \pi^{+} \pi^{-} K^{-}$, for $k^{2}/m_{B}^{2}=0.3(0.5)$ and for  model $(1)$.   The solid (dotted) line at
 $\sin \delta=+1$ corresponds the case    $\tilde{\Pi}_{\rho \omega}=(-3500;-300)$, where     
 $\rho - \omega$ mixing is included. The  dot-dashed (dot-dot-dashed) line corresponds to 
 $\tilde{\Pi}_{\rho \omega}=(0;0)$, where   $\rho - \omega$ mixing is not included. }
\item{ Fig.~\ref{fig12}  The ratio of penguin to tree amplitudes, $r$,  as a function of $N_{c}^{eff}$,
for ${\Bar B}^{0} \rightarrow
 \pi^{+} \pi^{-} {\Bar K}^{0}$,   for $k^{2}/m_{B}^{2}=0.3(0.5)$, for limiting values of the CKM matrix elements 
$(\rho,\eta)$ max(min), for  $\tilde{\Pi}_{\rho \omega}=(-3500;-300)(0,0)$, (i.e. with(without) $\rho - \omega$
  mixing) and for   model $(1)$. Figure 12a (left): for $\tilde{\Pi}_{\rho \omega}=(0;0)$,  solid line (dotted line)
 for
 $k^{2}/m_{B}^{2}=0.3$ and $(\rho,\eta)$ max(min). Dot-dashed line (dot-dot-dashed line) for  $k^{2}/m_{B}^{2}=0.5$
 and $(\rho,\eta)$ max(min). Figure 12b (right): same caption but for  $\tilde{\Pi}_{\rho \omega}=(-3500;-300)$.}
\item{ Fig.~\ref{fig13}  The ratio of penguin to tree amplitudes, $r$, for $B^{-} \rightarrow
 \pi^{+} \pi^{-} K^{-}$.  Same caption for 
Figure 13a (left) and Figure 13b (right) as in  Fig.~\ref{fig12}.}
\item{ Fig.~\ref{fig14}  Branching ratio for $B^{\pm} \rightarrow \rho^{0}  K^{\pm}$, for models  $1(2)$,
 $k^{2}/m_{B}^{2}=0.3$ and limiting values of the  CKM matrix elements. Solid line (dotted line) for  model $(1)$
 and max (min) CKM matrix elements. Dot-dashed line (dot-dot-dashed line) for   model $(2)$ and max (min) CKM 
matrix elements.}
\item{ Fig.~\ref{fig15} Branching ratio for $B^{\pm} \rightarrow \rho^{\pm}K^{0}$, for models  $1(2)$, 
$k^{2}/m_{B}^{2}=0.3$
 and limiting values of the  CKM matrix elements. Solid line (dotted line) for  model $(1)$ and max (min) CKM 
matrix elements. Dot-dashed line (dot-dot-dashed line) for  model $(2)$ and max (min) CKM matrix elements.}
\item{ Fig.~\ref{fig16} Branching ratio for $B^{0} \rightarrow \rho^{\pm}K^{\mp}$, for models  $1(2)$, 
$k^{2}/m_{B}^{2}=0.3$
 and limiting values of the  CKM matrix elements. Solid line (dotted line) for  model $(1)$ and max (min) CKM 
matrix elements. Dot-dashed line (dot-dot-dashed line) for  model $(2)$ and max (min) CKM matrix elements.}
\item{ Fig.~\ref{fig17} Branching ratio for $B^{0} \rightarrow \rho^{0}K^{0}$, for models  $1(2)$, 
$k^{2}/m_{B}^{2}=0.3$
 and limiting values of the  CKM matrix elements. Solid line (dotted line) for  model $(1)$ and max (min) CKM 
matrix elements. Dot-dashed line (dot-dot-dashed line) for  model $(2)$ and max (min) CKM matrix elements.}
\item{ Fig.~\ref{fig18} Branching ratio for $B^{\pm} \rightarrow \omega K^{\pm}$, for models  $1(2)$, $k^{2}
/m_{B}^{2}=0.3$
 and limiting values of the  CKM matrix elements. Solid line (dotted  line) for  model $(1)$ and max (min) CKM 
matrix elements. Dot-dashed line (dot-dot-dashed line) for  model $(2)$ and max (min) CKM matrix elements.}
\item{ Fig.~\ref{fig19} The ratio of the two $\rho K$ branching ratios versus $N_{c}^{eff}$ 
 for models $1(2)$ and for limiting values of the CKM matrix elements: solid line (dotted line) for   model $(1)$
 with max (min) CKM matrix elements.  Dot-dashed line (dot-dot-dashed line) for   model $(2)$ with max (min) 
CKM matrix elements.}
\end{itemize}
\newpage
%
\section*{Table  captions}
%
%
\begin{itemize}
\item{ Table~\ref{tab1} Wilson coefficients to the next-leading order.}
\item{ Table~\ref{tab2} Effective Wilson coefficients related to the tree operators, 
 electroweak and QCD-penguin operators.}
\item{ Table~\ref{tab4} Values of the CKM unitarity triangle  for limiting values of the CKM matrix elements.}
\item{ Table~\ref{tab5} Form factor  values for  $B \rightarrow \rho$ and $B \rightarrow  K$ at $q^{2}=0$.}
\item{ Table~\ref{tab7} Maximum  $CP$  violating asymmetry $a_{max}(\%)$ for ${\Bar B}^{0} \rightarrow \pi^{+} \pi^{-}
 {\Bar K}^{0}$, 
for all models, limiting values of the  CKM matrix elements (upper and lower limit), and for
 $k^{2}/m_{b}^{2}=0.3(0.5)$. }
\item{ Table~\ref{tab8}  Maximum  $CP$  violating asymmetry $a_{max}(\%)$ for $B^{-} \rightarrow \pi^{+} \pi^{-} K^{-}$, 
for all models, limiting values of the  CKM matrix elements (upper and lower limit), and for
 $k^{2}/m_{b}^{2}=0.3(0.5)$.  }
\item{ Table~\ref{tab6} The measured branching ratios by  CLEO, BABAR and BELLE factories for $B$ decays 
into $\rho K$ ($10^{-6}$).}
\item{ Table~\ref{tab15} Best range of  $N_{c}^{eff}$ determined  for 
$k^{2}/m_{b}^{2}=0.3(0.5)$ and for $ B \rightarrow \rho K$ decays (upper). Also range of
$N_{c}^{eff}$ determined previously  for $ B \rightarrow \rho \pi$ decays~\cite{refa3} (updated). Finally
global range of  $N_{c}^{eff}$ from both $B$ decays (lower). }
\end{itemize}
\clearpage
\clearpage
\clearpage
\clearpage
\clearpage
\clearpage
\clearpage
\clearpage
\clearpage
\clearpage
\renewcommand\baselinestretch{1.00}
%
%
\begin{table}
\begin{center}
\begin{tabular}{cccc} 
\hline 
\hline    
\multicolumn{4}{c}{$C_{i}(\mu)$ for $\mu= 5$ GeV} \\
\hline 
\hline    
 & $C_{1} $   & $-0.3125$   \\
 &  $C_{2} $   & $+1.1502$   \\
\cline{1-4}
$C_{3} $   & $+0.0174$   & $C_{5} $  & $+0.0104$ \\
$C_{4} $   & $+0.0373$   & $C_{6} $  & $-0.0459$ \\
\hline
$C_{7} $   & $-1.050 \times 10^{-5}$   & $C_{9} $  & $-0.0101$                \\
$C_{8} $   & $+3.839 \times 10^{-4}$   & $C_{10}$  & $+1.959 \times 10^{-3}$  \\
\hline
\hline
\end{tabular}
\end{center}
\caption{Wilson coefficients to the next-leading order (see the reference in text). }
\label{tab1}
\end{table}
%
%
\begin{table}
\begin{center}
\begin{tabular}{ccc} \hline \hline    
   $C_{i}^{\prime}$          & $q^{2}/m_{b}^{2}=0.3$   & $q^{2}/m_{b}^{2}=0.5$ \\
\hline
\hline
$C_{1}^{\prime} $   &   $-0.3125$                                     &$-0.3125 $  \\
$C_{2}^{\prime} $   & $+1.1502$                                        &$+1.1502$    \\
\hline
$C_{3}^{\prime} $   & $+2.433 \times 10^{-2} + 1.543 \times 10^{-3}i$  &$+2.120 \times 10^{-2} + 2.174 
\times 10^{-3}i$\\
$C_{4}^{\prime} $   & $-5.808 \times 10^{-2} -4.628 \times 10^{-3}i$  &$-4.869 \times 10^{-2} -1.552 
\times 10^{-2}i$\\
$C_{5}^{\prime} $   & $+1.733 \times 10^{-2}+ 1.543 \times 10^{-3}i$   &$+1.420 \times 10^{-2} + 5.174 
\times 10^{-3}i$\\
$C_{6}^{\prime} $   & $-6.668 \times 10^{-2}- 4.628 \times 10^{-3}i$  &$-5.729  \times 10^{-2}- 1.552 
\times 10^{-2}i$\\
\hline
$C_{7}^{\prime} $   & $-1.435 \times 10^{-4} -2.963 \times 10^{-5}i$  &$-8.340 \times 10^{-5} -9.938 
\times 10^{-5}i$\\
$C_{8}^{\prime} $   & $+3.839 \times 10^{-4}$                          &$ +3.839 \times 10^{-4} $\\
$C_{9}^{\prime} $   & $-1.023 \times 10^{-2} -2.963 \times 10^{-5}i$  &$-1.017 \times 10^{-2} -9.938 
\times 10^{-5}i$\\
$C_{10}^{\prime} $  & $+1.959 \times 10^{-3}$                          &$+1.959 \times 10^{-3}$\\
\hline
\hline
\end{tabular}
\end{center}
\caption{ Effective Wilson coefficients related to the tree operators, 
 electroweak and QCD penguin operators (see the reference  in text). }
\label{tab2}
\end{table}
%
%
\begin{table}
\begin{center}
\begin{tabular}{cccc} \hline \hline    
                          &      $\alpha$       &      $\beta$       &       $\gamma$   \\
\hline
\hline
 $(\rho_{min},\eta_{min})$     &       $ 104^{o}47$    &     $ 19^{o}32$    &       $  56^{o}21$    \\
\hline
 $(\rho_{min},\eta_{max})$  &       $ 93^{o}13$    &     $ 24^{o}31$ &      $ 62^{o}56$              \\
\hline  
 $(\rho_{max},\eta_{min})$  &       $ 112^{o}14$    &     $ 21^{o}20$ &      $ 46^{o}66$              \\       
\hline  
 $(\rho_{max},\eta_{max})$ &         $ 99^{o}66$    &     $ 26^{o}56$ &      $ 53^{o}78$              \\       
\hline  
\hline
\end{tabular}
\end{center}
\caption{Values of the CKM unitarity triangle  for limiting values of the CKM matrix elements. }
\label{tab4}
\end{table}
%
%
\newpage
\begin{table}[htp]
\begin{center}
\begin{tabular}{ccccccc} \hline \hline    
         &      $h_{A_{0}}$     & $h_{1}$    &   $m_{A_{0}}$  &   $m_{1}$  & $d_{0}(d_{1})$   &   $b_{0}(b_{1})$  \\ 
\hline
\hline
 model $(1)$ & 0.280 & 0.360 & 5.27 & 5.41 &              &               \\
\hline
 model $(2)$ & 0.340 & 0.762 & 5.27 & 5.41 &              &               \\
\hline
 model $(3)$ & 0.280 & 0.360 & 5.27 & 5.41 &              &               \\ 
\hline
 model $(4)$ & 0.340 & 0.762 & 5.27 & 5.41 &              &               \\ 
\hline
 model $(5)$ & 0.372 & 0.341 &      &      & 1.400(0.410) & 0.437(-0.361) \\ 
\hline
\hline
\end{tabular}
\end{center}
\caption{Form factor  values for  $B \rightarrow \rho$ and $ B \rightarrow K$ at $q^{2}=0$ (see the reference
 in text). }
\label{tab5}
\end{table}
%
%
%
\begin{table}[hpb]
\begin{center}
\begin{tabular}{ccc} \hline \hline    
                      & $N_{cmin}^{eff}=0.66(0.61)$   & $N_{cmax}^{eff}=2.84(2.82)$ \\
\hline
\hline
model $(1)$                                                                         \\
\hline
\hline
$ \rho_{max},\eta_{max}$   &        32(46)    &      -14(-16)                   \\
$ \rho_{min},\eta_{min}$   &        25(33)    &      -19(-22)                   \\
\hline                                  
\hline
model $(2)$                                                                        \\               
\hline
\hline
$ \rho_{max},\eta_{max}$   &        32(41)    &      -6(-7)                   \\
$ \rho_{min},\eta_{min}$   &        27(30)    &      -9(-10)                   \\
\hline  
\hline  
model $(3)$                                                                         \\
\hline
\hline
$ \rho_{max},\eta_{max}$   &        32(45)    &      -14(-16)                   \\ 
$ \rho_{min},\eta_{min}$   &        25(33)    &      -20(-23)                   \\
\hline  
\hline 
model $(4)$                                                                         \\
\hline
\hline
$ \rho_{max},\eta_{max}$   &        32(41)    &      -6(-7)                   \\ 
$ \rho_{min},\eta_{min}$   &        27(30)    &      -9(-10)                   \\ 
\hline  
\hline
model $(5)$                                                                         \\
\hline
\hline
$ \rho_{max},\eta_{max}$   &        37(55)    &      -15(-17)                   \\
$ \rho_{min},\eta_{min}$   &        26(40)    &      -19(-24)                   \\
\hline
\hline
\end{tabular}
\end{center}
\caption{Maximum  $CP$  violating asymmetry $a_{max}(\%)$ for ${\bar B}^{0} \rightarrow \pi^{+} \pi^{-} {\bar K}^{0}$, 
for all models, limiting values (upper and lower) of the  CKM matrix elements, and for
 $k^{2}/m_{b}^{2}=0.3(0.5)$. }
\label{tab7}
\end{table}
%
%
%
\begin{table}[hpb]
\begin{center}
\begin{tabular}{ccc} \hline \hline    
                      & $N_{cmin}^{eff}=0.66(0.61)$   & $N_{cmax}^{eff}=2.84(2.82)$ \\
\hline
\hline
model $(1)$                                                                         \\
\hline
\hline
$ \rho_{max},\eta_{max}$   &        47(45)    &      -15(-17)                   \\
$ \rho_{min},\eta_{min}$   &        34(35)    &      -21(-23)                   \\
\hline                                  
\hline
model $(2)$                                                                        \\               
\hline
\hline
$ \rho_{max},\eta_{max}$   &        45(41)    &      -11(-13)                   \\
$ \rho_{min},\eta_{min}$   &        33(32)    &      -17(-18)                   \\
\hline  
\hline  
model $(3)$                                                                         \\
\hline
\hline
$ \rho_{max},\eta_{max}$   &        47(44)    &      -15(-17)                   \\ 
$ \rho_{min},\eta_{min}$   &        34(35)    &      -20(-23)                   \\
\hline  
\hline 
model $(4)$                                                                         \\
\hline
\hline
$ \rho_{max},\eta_{max}$   &        45(42)    &      -12(-13)                   \\ 
$ \rho_{min},\eta_{min}$   &        33(32)    &      -17(-18)                   \\ 
\hline  
\hline
model $(5)$                                                                         \\
\hline
\hline
$ \rho_{max},\eta_{max}$   &        49(46)    &      -17(-19)                   \\
$ \rho_{min},\eta_{min}$   &        36(35)    &      -22(-25)                   \\
\hline
\hline
\end{tabular}
\end{center}
\caption{Maximum  $CP$  violating asymmetry $a_{max}(\%)$ for $B^{-} \rightarrow \pi^{+} \pi^{-} K^{-}$, 
for all models, limiting values of the  CKM matrix elements (upper and lower limit), and 
for $k^{2}/m_{b}^{2}=0.3(0.5)$. }
\label{tab8}
\end{table}
%
%
\newpage
\begin{table}[htp]
\begin{center}
\begin{tabular}{cccc} \hline \hline    
                    &             CLEO             &        BABAR               &  BELLE            \\ 
\hline
\hline
$\rho^{0} K^{\pm}$  & ${8.46^{+4.0}_{-3.4} \pm 1.8}^{\bullet}$ ${(\leq 17)}^{\P}$  &   ${10 \pm 6 \pm 2}^{\star} $
  ${(\leq 29)}^{\P}$ & ${\leq 13.5}^{\P}$ \\
\hline
$\rho^{\pm} K^{0}$  &        $-$              &        $-$           &      ${\leq 23.6}^{\P}$        \\
\hline
$\rho^{\pm} K^{\mp}$ & ${16.0^{+7.6}_{-6.4} \pm 2.8}^{\bullet} $ ${(\leq 32)}^{\P}$&        $-$           &  
 ${15.8^{+5.1 \; +1.7}_{-4.6 \; -3.0}}^{\star} $      \\ 
\hline
$\rho^{0} K^{0}$    &             $-$               &       $-$             &       $-$                   \\ 
\hline
$\frac{BR(\rho^{\pm} K^{\mp})}{BR(\rho^{0} K^{\pm})}$ & $1.89 \pm 1.41$   &   $-$    &    $-$     \\         
\hline
$ \omega K^{\pm}$ &  ${3.2^{+2.4}_{-1.9} \pm 0.8}^{\bullet}$ ${(\leq 7.9)}^{\P}$ & ${1.4^{+1.3}_{-1.0} \pm
 0.3}^{\star}$ & $ {9.2^{+2.6}_{-2.3} \pm 1.0}^{\star}$ \\
\hline
\hline
\end{tabular}
\end{center}
\caption{The measured branching ratios by  CLEO, BABAR and BELLE factories for $B$ decays into $\rho K$ 
($10^{-6}$) (see the reference in text). ${\rm Experimental \; data}^{\star}$, 
${\rm fit}^{\bullet}$ and  ${\rm upper \; limit}^{\P}$.}
\label{tab6}
\end{table}
%
%
%
\begin{table}[p]
\begin{center}
\begin{tabular}{cc} \hline \hline   
    $B \rightarrow \rho K  $  &   $\left\{N_{c}^{eff}\right\}$     \\
\hline \hline 
model $(1)$               & 0.66;2.68(0.61;2.68)           \\
model $(2)$               & 1.17;2.84(1.09;2.82)       \\   
\hline 
\hline
maximum range           & 0.66;2.84(0.61;2.82)               \\
minimum range           & 1.17;2.68(1.09;2.68)             \\
\hline
\hline
         &                      \\
\hline
\hline
    $B \rightarrow \rho \pi$    &   $\left\{N_{c}^{eff}\right\}$     \\
\hline \hline 
model $(2)$               & 1.09;1.63(1.12;1.77)             \\
model $(4)$               & 1.10;1.68(1.11;1.80)         \\   

\hline
\hline
maximum range           & 1.09;1.68(1.11;1.80)                 \\
minimum range           & 1.10;1.63(1.12;1.77)              \\
\hline
\hline
         &                      \\
\hline
\hline
     global range    &      $\left\{N_{c}^{eff}\right\}$                   \\
\hline
\hline
global maximum range           & 0.66;2.84(0.61;2.82)                       \\
global minimum range           & 1.17;1.63(1.12;1.77)                     \\
\hline
\hline
\end{tabular}
\end{center}
\caption{Best range of  $N_{c}^{eff}$ determined  for 
$k^{2}/m_{b}^{2}=0.3(0.5)$ and for $ B \rightarrow \rho K$ decays (upper). Also range of
$N_{c}^{eff}$ determined previously  for $ B \rightarrow \rho \pi$ decays~\cite{refa3} (updated). Finally
global range of  $N_{c}^{eff}$ from both $B$ decays (lower).}
\label{tab15}
\end{table}
%
%
\end{document}